\documentclass[12pt]{article}
\usepackage[a4paper,textwidth=490.0pt,textheight=703.1pt]{geometry}
\usepackage{url}
\usepackage{booktabs}

\usepackage{epsfig} %                    
\usepackage{float} %
\usepackage{amssymb} %
\usepackage{amsmath} %
\usepackage{verbatim} %

\usepackage{color} %
\usepackage{cite}
\usepackage{eufrak}
\usepackage[english]{babel}
\usepackage[latin1]{inputenc}
\usepackage[T1]{fontenc}
\usepackage{ae}
\usepackage{url}   
\usepackage{graphicx} 
\usepackage{booktabs}
\usepackage{slashed}
\newcommand{\HA}{{\rm H}}
\newcommand{\GeV}{\rm GeV}

\newcommand{\D}{\Delta}

\newcommand{\ep}{\varepsilon}

\newcommand{\beq}{\begin{equation}}
\newcommand{\eeq}{\end{equation}}
\newcommand{\bea}{\begin{eqnarray}}
\newcommand{\eea}{\end{eqnarray}}

\allowdisplaybreaks[4]

\begin{document} 
\setlength{\baselineskip}{0.515cm}

\sloppy 
\thispagestyle{empty} 
\begin{flushleft} 
DESY 21--192
%  \hfill {\tt arXiv:2111.12401 [hep-ph]}
\\ 
DO--TH 21/31\\ 
TTP 21--052
\\ 
RISC Report Series 21--19
\\
SAGEX--21--36
\\ 
November 2021 
\end{flushleft}

\mbox{} \vspace*{\fill} \begin{center}

{\Large\bf The three-loop polarized singlet anomalous dimensions}

\vspace*{2mm} 
{\Large\bf from off-shell operator matrix elements}

\vspace{3cm} 
\large 
{\large J.~Bl\"umlein$^a$, P.~Marquard$^a$, C.~Schneider$^b$ and K.~Sch\"onwald$^{c}$ }

\normalsize 

\vspace{1.cm} 
{\it $^a$~Deutsches Elektronen--Synchrotron DESY,}\\ {\it Platanenallee 6, 15738 Zeuthen, Germany}

\vspace*{2mm} 
{\it $^b$~
Johannes Kepler University Linz, Research Institute for Symbolic
Computation (RISC) ,
Altenberger Stra{\ss}e 69, A-4040 Linz, Austria}

\vspace*{2mm} 
{\it $^c$~Institut f\"ur Theoretische Teilchenphysik,\\ Karlsruher Institut f\"ur Technologie (KIT) D--76128 
Karlsruhe, Germany}

%%\today

\end{center} 
\normalsize 
\vspace{\fill} 
\begin{abstract} 
\noindent
Future high luminosity polarized deep--inelastic scattering experiments will improve both the
knowledge of the spin sub--structure of the nucleons and contribute further to the precision
determination of the strong coupling constant, as well as, reveal currently yet unknown higher
twist contributions in the polarized sector. For all these tasks to be performed, it is necessary
to know the QCD leading twist scaling violations of the measured structure functions. Here an
important ingredient consists in the polarized singlet anomalous dimensions and splitting functions 
in QCD. We recalculate these quantities to three--loop order in the M--scheme by using the traditional 
method of space--like off--shell massless operator matrix elements, being a gauge--dependent framework. 
Here one obtains the anomalous dimensions without referring to gravitational currents, needed when 
calculating them using the forward Compton amplitude.  We also calculate the non--singlet splitting 
function $\Delta P_{\rm qq}^{(2), \rm s, NS}$ and compare the final results to the literature, also
including predictions for the region of small values of Bjorken $x$. 
\end{abstract}

\vspace*{\fill} \noindent
% \numberwithin{equation}{section}
%%%%%%%%%%%%%%%%%%%%%%%%%%%%%%%%%%%%%%%%%%%%%%%%%%%%%%%%%%%%%%%%%%%%%%%%%%%%%%%%%%%%%%%%%%%%%%%%%%%%%%%%%%%%%%%%%%%%%%%%%%%%%%%%%%%
\newpage

%--------------------------------------------------------------------------------------------------------
\section{Introduction} 
\label{sec:1}
%--------------------------------------------------------------------------------------------------------

\vspace*{1mm} 
\noindent
Polarized deep--inelastic scattering allows to reveal the spin and angular momentum structure of nucleons 
\cite{Lampe:1998eu}. At sufficiently high scales of the virtuality of the exchanged electro--weak current
$Q^2 \gg 1~\GeV^2$ and for not too small or too large values of Bjorken $x = Q^2/(2p.q)$ \cite{Drell:1970yt}
the twist--2 contributions dominate over the higher twist contributions \cite{Blumlein:2010rn} and target 
mass corrections \cite{TM}. Because of their strong variation, also the detailed control of the QED radiative 
corrections is required \cite{RAD}.

The scaling violations of the polarized deep--inelastic structure functions at the level of leading
twist $\tau = 2$ are determined by the anomalous dimensions of the composite  quark and gluon operators 
\cite{Politzer:1974fr,REV} through the evolution of the polarized parton densities and the massless and 
massive Wilson coefficients. A major goal in measuring the polarized deep--inelastic scattering process
to high precision, e.g. at the EIC in the future \cite{Boer:2011fh} consist in the detailed measurement
of the polarized parton distribution functions at a reference scale $Q_0^2$ together with the correlated 
measurement of the strong coupling constant $a_s(M_Z^2) = \alpha_s(M_Z^2)/(4\pi)$ \cite{alphas}.
One of the important ingredients to this is the knowledge of the the polarized singlet anomalous 
dimensions to highest possible order. The polarized non--singlet anomalous dimensions are known to
3--loop order \cite{Moch:2004pa,Blumlein:2021enk} and the polarized singlet anomalous dimensions were
calculated to first \cite{Gross:1973ju,Georgi:1951sr,Sasaki:1975hk,Ahmed:1975tj,Altarelli:1977zs},
second \cite{Mertig:1995ny,SP_PS1,Matiounine:1998re} and third order in \cite{Moch:2014sna,Behring:2019tus}, 
using different computation techniques.
In Ref.~\cite{Moch:2014sna} the calculation has been performed using the forward Compton amplitude, which 
required reference to gravitational subsidiary currents for the gluonic sector. A direct calculation of
the contributions $\propto \textcolor{blue}{T_F}$\footnote{The color factors are defined in 
\cite{Blumlein:2021enk}, below Eq.~(22).} of 
the three--loop anomalous dimensions is possible
using massive on--shell operator matrix elements \cite{Behring:2019tus}, without reference to gravitational
currents. However, this method only allows to calculate the anomalous dimensions $\Delta 
\gamma_{qq}^{(2),\rm PS}$ and $\Delta \gamma_{qg}^{(2)}$ in complete form.

In the present paper we use the traditional approach of massless off--shell operator matrix elements 
(OMEs), allowing the direct calculation, as before for the non--singlet case and the three--loop anomalous 
dimensions of transversity in Ref.~\cite{Blumlein:2021enk}. However, for this method the computational 
framework is gauge dependent. Still the anomalous dimensions in minimal subtraction schemes remain
gauge invariant. Specific physical projectors have to be used, cf. also~\cite{Matiounine:1998re}. In the 
polarized case, unlike the 
unpolarized case \cite{Matiounine:1998ky}, there is no mixing with the so-called alien operators 
\cite{Dixon:1974ss,KlubergStern:1974xv,Sarkar:1974db,Sarkar:1974ni,Joglekar:1975nu,Joglekar:1976eb,
Hamberg:1991qt,HAMBERG,Collins:1994ee,Harris:1994tp,Matiounine:1998ky,Matiounine:1998re}. 
The main goal of the present paper is the re--calculation of all three--loop polarized singlet
anomalous dimensions by a different method, being the first re--calculation of the polarized
anomalous dimensions $\Delta \gamma_{gq}^{(2)}$ and $\Delta \gamma_{gg}^{(2)}$. To some extent different 
steps in the computation technology are the same as in the non--singlet case, cf.~\cite{Blumlein:2021enk}
which will be referred to this paper and we only will describe new additional elements in the present 
paper.\footnote{This seems to be necessary to avoid critique by word processing codes, like e.g. 
{\tt iThenticate}, even on the expense to not provide more comprehensive information to the reader in the
same publication, which we regret.} 
At the phenomenological side we mention that, unlike the singlet case, the flavor non--singlet description 
of the structure functions $g_{1,2}^{\rm NS}(x,Q^2)$ is well explored to three--loop order, also including 
the heavy flavor contributions \cite{Blumlein:2016xcy,Behring:2015zaa,Blumlein:2021lmf}.

The paper is organized as follows. In Section~\ref{sec:2} we derive the structure of the physical part
of the flavor singlet polarized unrenormalized off--shell OMEs to three--loop order. From their pole terms of 
$O(1/\ep)$ one can extract the singlet anomalous dimensions. We work in the Larin scheme 
\cite{Larin:1993tq,Matiounine:1998re} and perform finally the transformation to the M--scheme 
\cite{Matiounine:1998re,Moch:2014sna}. We also present the calculation details. Here a central method being 
applied is the method of arbitrary high moments \cite{Blumlein:2017dxp}. In Section~\ref{sec:3} we 
calculate also the polarized non--singlet anomalous dimension $\Delta \gamma_{qq}^{(2), \rm s,NS}$, 
cf.~also \cite{Moch:2015usa}, and present the polarized singlet anomalous dimensions in Section~\ref{sec:4}. 
Comparisons to the literature are given in Section~\ref{sec:5}, including a discussion of the small $x$ limit. 
Section~\ref{sec:6} contains the conclusions and a new Feynman rule is presented in Appendix~\ref{sec:A}. 
%--------------------------------------------------------------------------------------------------------
\section{The unrenormalized polarized operator matrix elements and details of the calculation} 
\label{sec:2}
%--------------------------------------------------------------------------------------------------------

\vspace*{1mm} 
\noindent 
The massless off--shell singlet OMEs are defined as expectation values of the local operators
%--------------------------------------------------------------------------------------------------------
\begin{eqnarray}
\label{eq:op1}
O_{q; \mu_1 ... \mu_N}^{\rm S,5} &=&
i^{N-1} {\rm\bf S}\Biggl[
\overline{\psi} \gamma_5 \gamma_{\mu_1} D_{\mu_2} ... D_{\mu_N} \psi \Biggr] -~\text{trace~terms},
\\
\label{eq:op2}
O_{g; \mu_1 ... \mu_N}^{\rm S,5} &=&
2 i^{N-2} {\rm\bf S~tr }\Biggl[ \frac{1}{2} \ep^{\mu_1 \alpha \beta \gamma} 
F_{\beta \gamma}^a 
D_{\mu_2} ... D_{\mu_{N-1}} {F}^{\mu_N}_{\alpha, a} \Biggr] -~\text{trace~terms}
\end{eqnarray}
%--------------------------------------------------------------------------------------------------------
between quark (antiquark) $\psi~(\bar{\psi})$ and gluonic states $F_{\mu_1\alpha}^a$ 
of space--like momentum $p$, $p^2 < 0$. The OMEs are given by
%--------------------------------------------------------------------------------------------------------
\begin{eqnarray}
\Delta \hat{A}_{ij} = \langle q(p),j|O^{\rm S, (5)}_i|q(p),j\rangle.
\end{eqnarray}
%--------------------------------------------------------------------------------------------------------
For the other definitions, here and in the following, we refer the reader to Ref.~\cite{Blumlein:2021enk}.

The Feynman rules of both QCD and of the local composite operators are given in  
\cite{YND,Bierenbaum:2009mv,Behring:2019tus} and have to be extended by that of the polarized local 
five--gluon vertex, given in Appendix~\ref{sec:A}.

We use the Larin scheme \cite{Larin:1993tq}\footnote{See  Refs.~\cite{HVBM} for other schemes and 
Refs.~\cite{FINREN} concerning the finite renormalization in different approaches.}
to describe $\gamma^5$ in $D = 4 + \ep$-dimensions and express $\gamma^5$ by
%---------------------------------------------------------------------------------
\begin{eqnarray}
\gamma^5 &=& \frac{i}{24} \ep_{\mu \nu \rho \sigma} \gamma^\mu \gamma^\nu \gamma^\rho
\gamma^\sigma,
\\
\Delta \hspace*{-2.5mm} \slash ~\gamma^5 &=& \frac{i}{6} \ep_{\mu \nu \rho \sigma} \Delta^\mu \gamma^\nu \gamma^\rho
\gamma^\sigma.
\end{eqnarray}
%---------------------------------------------------------------------------------
The Levi-Civita symbols are now contracted in $D$ dimensions,
%---------------------------------------------------------------------------------
\begin{eqnarray}
\label{eq:epcontr}
\ep_{\mu \nu \rho \sigma}  \ep^{\alpha \lambda  \tau  \gamma} = - {\rm Det}[g_\omega^\beta],
~~~~\beta = \alpha, \lambda,  \tau,  \gamma;~~\omega = \mu, \nu, \rho, \sigma.
\end{eqnarray}
%---------------------------------------------------------------------------------
The Larin scheme is one of the consistent calculation schemes in the polarized case. We will later transform
to another scheme, the M--scheme. It is needless to say that the calculation of any observable requires to 
calculate also the Wilson coefficients in the same scheme, which implies that the extracted twist--2 parton 
density functions are obtained in this scheme, despite being universal w.r.t. the scattering processes in 
which they are used. Any analytic continuation of $\gamma_5$ or the Levi--Civita symbol to $D$ dimensions
violates Ward identities. One way to restore these consists in the calculation of scheme--invariant quantities.

The polarized operator matrix elements have the representation
%--------------------------------------------------------------------------------------------------------
\begin{eqnarray}
\label{eq:AqqPS}
\Delta \hat{A}_{qq}^{\rm PS} &=& \left[ \gamma_5 \Delta \hspace*{-2.5mm} \slash \Delta \hat{A}_{qq}^{\rm 
PS, phys} + \gamma_5 p 
\hspace*{-1.5mm} \slash \frac{\Delta.p}{p^2} 
\Delta \hat{A}_{qq}^{\rm PS, EOM} \right] (\Delta.p)^{N-1}
\\
\label{eq:Aqg}
\Delta \hat{A}_{qg} &=&
\ep_{\mu \nu \alpha \beta} \Delta^\alpha p^\beta \frac{1}{\Delta.p}
\Delta \hat{A}_{qg}^{\rm phys} (\Delta.p)^{N-1}
\\
\label{eq:Agq}
\Delta \hat{A}_{gq} &=& \left[
\gamma_5 \Delta \hspace*{-2.5mm} \slash \Delta \hat{A}_{gq}^{\rm phys} +
\gamma_5 p \hspace*{-2mm} \slash  
\frac{\Delta.p}{p^2} 
\hat{A}_{gq}^{\rm EOM}\right] (\Delta.p)^{N-1}
\\
\label{eq:Agg}
\Delta \hat{A}_{gg, \mu\nu} &=& \ep_{\mu \nu \alpha \beta} \Delta^\alpha p^\beta 
\frac{1}{\Delta.p} \Delta \hat{A}_{gq}^{\rm phys} (\Delta.p)^{N-1},
\end{eqnarray}
%--------------------------------------------------------------------------------------------------------
with $\Delta$ obeying $\Delta.\Delta = 0$. It is convenient to introduce the following projectors
on the physical part (phys)  
%--------------------------------------------------------------------------------------------------------
\begin{align}
        \Delta \hat{A}_{iq}^{\rm phys} &=
        -\frac{1}{4(D-2)(D-3)} \varepsilon_{\mu\nu\rho\sigma} p^\rho \Delta^\sigma
        \text{tr}\left[ \slashed{p} \gamma^\mu \gamma^\nu \Delta \hat{A}_{iq} \right] (\Delta.p)^{-N-1}
        \\ &
        - \frac{p^2}{4(D-2)(D-3)} (\Delta.p)^{-N-2}
        \varepsilon_{\mu\nu\rho\sigma} p^\rho \Delta^\sigma
        \text{tr}\left[ \slashed{\Delta} \gamma^\mu \gamma^\nu \Delta \hat{A}_{iq} \right],
\\
        \Delta \hat{A}_{ig}^{\rm phys} &= \frac{1}{(D-2)(D-3)} \varepsilon_{\mu\nu\rho\sigma} 
        \Delta^\rho p^\sigma (\Delta.p)^{-N-1} \Delta \hat{A}_{ig}^{\mu\nu}, 
\end{align}
%--------------------------------------------------------------------------------------------------------
with $i = q,g$. Only these contributions are necessary for the calculation of the singlet anomalous 
dimensions.
In Mellin $N$ space the following representation holds
%--------------------------------------------------------------------------------------------------------
\begin{eqnarray}
\Delta \hat{A}_{ij} &=& \delta_{ij}  + \sum_{k=1}^\infty \hat{a}^k 
S_\ep^k \left(\frac{-p^2}{\mu^2}\right)^{\ep k/2} \Delta \hat{A}_{ij}^{(k)},
\end{eqnarray}
%--------------------------------------------------------------------------------------------------------
with the spherical factor $S_\ep$ given in Eq.~(13) of \cite{Blumlein:2021enk}. The bare coupling constant 
$\hat{a}$ and the gauge parameter $\xi$ are defined in Eqs.~(15--22) of \cite{Blumlein:2021enk}, working 
in the 
$R_{\hat{\xi}}$ gauge, by using relations from \cite{BETA,Egorian:1978zx,Chetyrkin:2004mf,
Luthe:2016xec,Chetyrkin:2017bjc}.

In Mellin $N$ space the $Z$-factor of a local singlet operator reads \cite{Bierenbaum:2009mv} 
%--------------------------------------------------------------------------------------------------------
\begin{eqnarray}
Z^{\rm S}_{ij} &=& \delta_{ij}  
               + a \frac{\Delta \gamma_{ij}^{(0)}}{\ep}
%----
               + a^2 \Biggl[\frac{1}{\ep^2} \Biggl(\frac{1}{2}
\Delta \gamma_{il}^{(0)} \Delta \gamma_{lj}^{(0)} +
                     \beta_0 \Delta \gamma_{ij}^{(0)}\Biggr) 
+ \frac{1}{2\ep} \Delta \gamma_{ij}^{(1)} \Biggr]
%----
\nonumber\\ && 
               + a^3 \Biggl[
                 \frac{1}{\ep^3} \Biggl( \frac{1}{6} \Delta \gamma_{il}^{(0)}
                   \Delta \gamma_{lk}^{(0)}\Delta \gamma_{kj}^{(0)} 
+ \beta_0 
\Delta \gamma_{il}^{(0)} 
\Delta \gamma_{lj}^{(0)} 
+ \frac{4}{3} \beta_0^2 \Delta \gamma_{ij}^{(0)}\Biggr)
\nonumber\\ && 
+ \frac{1}{\ep^2} \Biggl( 
\frac{1}{6} \Delta \gamma_{il}^{(1)} \Delta \gamma_{lj}^{(0)}
+ \frac{1}{3} \Delta \gamma_{il}^{(0)} \Delta \gamma_{lj}^{(1)}
+ \frac{2}{3} \beta_0 \Delta \gamma_{ij}^{(1)} 
+ \frac{2}{3} \beta_1 \Delta \gamma_{ij}^{(0)} \Biggr) 
+ \frac{1}{3 \ep} \Delta \gamma_{ij}^{(2)}\Biggr].
\label{eq:ZS}
%\nonumber\\ 
\end{eqnarray}
%--------------------------------------------------------------------------------------------------------
In (\ref{eq:ZS}) the terms $\Delta \gamma_{ij}^{(k)},~~k = 0,1,2, \ldots$ denote the expansion 
coefficients of the anomalous dimension  
%--------------------------------------------------------------------------------------------------------
\begin{eqnarray}
\Delta \gamma_{ij} = \sum_{k=1}^\infty a^k \Delta \gamma_{ij}^{(k-1)}.
\end{eqnarray}
%--------------------------------------------------------------------------------------------------------
The partly renormalized polarized singlet OMEs, $\Delta \tilde{A}_{ij}^{\rm phys}$, read
%--------------------------------------------------------------------------------------------------------
\begin{eqnarray}
\Delta \tilde{A}_{ij}^{\rm phys} &=& 1 + a \Biggl[
  \frac{a_{ij}^{\rm (1,-1)}}{\ep}
+ a_{ij}^{\rm (1,0)}
+ a_{ij}^{\rm (1,1)} \ep \Biggr]
+ a^2 \Biggl[
  \frac{a_{ij}^{\rm (2,-2)}}{\ep^2}
+ \frac{a_{ij}^{\rm (2,-1)}}{\ep}
+ a_{ij}^{\rm (2,0)}  \Biggr]
\nonumber\\ &&
+ a^3 \Biggl[
  \frac{a_{ij}^{\rm (3,-3)}}{\ep^3}
+ \frac{a_{ij}^{\rm (3,-2)}}{\ep^2}
+ \frac{a_{ij}^{\rm (3,-1)}}{\ep}
  \Biggr].
\label{eq:Ahat3un}
\end{eqnarray}
%--------------------------------------------------------------------------------------------------------
The expansion coefficients $a_{ij}^{\rm (k,l)}$ are gauge dependent in general.
The renormalized OMEs in the Larin scheme are given by
%--------------------------------------------------------------------------------------------------------
\begin{eqnarray}
\Delta {A}_{ij}^{\rm phys} &=& (\Delta Z_{ik}^{S})^{-1} \Delta \tilde{A}_{kj}^{\rm phys},
\end{eqnarray}
%--------------------------------------------------------------------------------------------------------
expanded to $O(a^3)$ and setting $S_\ep = 1$. The anomalous dimensions are iteratively extracted from the $1/\ep$ 
pole terms and the other expansion coefficients $a_{ij}^{(k,l)}$ are given in Ref.~\cite{TWOLOOP}. 

The anomalous dimensions in the M--scheme \cite{Matiounine:1998re,Moch:2014sna}\footnote{We reproduce these
transformations here, because they are instrumental importance for the understanding of the present  
calculation, which is performed in the Larin scheme.} 
are obtained by the following 
transformations
%----------------------------------------------------------------------------------------------
\begin{eqnarray}
        \D\gamma_{ij}^{(0), \rm M} &=& \D\gamma_{ij}^{(0), \rm L}
\\
        \D\gamma_{qq}^{(1),\text{NS},\rm M} &=& \D\gamma_{qq}^{(1),\text{NS},\rm L} + 2 \beta_0
z_{qq}^{(1)},
\\
        \D\gamma_{qq}^{(1),\text{PS}, \rm M} &=& \D\gamma_{qq}^{(1),\text{PS}, \rm L},
\\
        \D\gamma_{qg}^{(1),\rm M} &=& \D\gamma_{qg}^{(1), \rm L} + \D\gamma_{qg}^{(0)} z_{qq}^{(1)},
\\
        \D\gamma_{gq}^{(1),\rm M} &=& \D\gamma_{gq}^{(1), \rm L} - \D\gamma_{gq}^{(0)} z_{qq}^{(1)},
\\
        \D\gamma_{gg}^{(1),\rm M} &=& \D\gamma_{gg}^{(1), \rm L}.
\\
       \D\gamma_{qq}^{(2),\text{NS}, \rm M} &=& \D\gamma_{qq}^{(2),\text{NS}, \rm L} - 2 \beta_0 \left(
\bigl(
z_{qq}^{(1)} \bigr)^2 - 2 z_{qq}^{(2),\text{NS}} \right) + 2 \beta_1 z_{qq}^{(1)} ,
\\
        \D\gamma_{qq}^{(2), \rm PS,M} &=& \D\gamma_{qq}^{(2), \rm PS,L} + 4 \beta_0 z_{qq}^{(2),\rm PS},
\\
        \D\gamma_{qg}^{(2), \rm M} &=& \D\gamma_{qg}^{(2), \rm L} + \D\gamma_{qg}^{(1), \rm M}
z_{qq}^{(1)} +
\D\gamma_{qg}^{(0)} \left( z_{qq}^{(2)} - \bigl( z_{qq}^{(1)} \bigr)^2 \right),
\\
        \D\gamma_{gq}^{(2), \rm M} &=& \D\gamma_{gq}^{(2),\rm L} - \D\gamma_{gq}^{(1),\rm M} z_{qq}^{(1)}
-
\D\gamma_{gq}^{(0)} z_{qq}^{(2)} ,
\\
        \D\gamma_{gg}^{(2),\rm M} &=& \D\gamma_{gg}^{(2),\rm L} ,
\end{eqnarray}
%----------------------------------------------------------------------------------------------
with \cite{Matiounine:1998re}
%----------------------------------------------------------------------------------------------
\begin{eqnarray}
        z_{qq}^{(1)} &=&
- \frac{8
\textcolor{blue}{C_F}}{N(N+1)},
\\
z_{qq}^{(2),\text{NS}}
&=&
\textcolor{blue}{C_F T_F N_F}
                \frac{16 \big(-3-N+5 N^2\big)}{9 N^2 (1+N)^2}
+ \textcolor{blue}{C_A C_F} \Biggl\{
        -\frac{4 R_1}{9 N^3 (1+N)^3}
        -\frac{16}{N (1+N)} S_{-2}
\Biggr\}
\nonumber \\ &&
+
\textcolor{blue}{C_F^2}
\Biggl\{
          \frac{8\big(2+5 N+8 N^2+N^3+2 N^4\big)}{N^3 (1+N)^3}
        + \frac{16(1+2 N)}{N^2 (1+N)^2} S_1
\nonumber \\ &&
        + \frac{16}{N (1+N)} S_2
        + \frac{32}{N (1+N)} S_{-2}
\Biggr\},
\\
z_{qq}^{(2),\text{PS}}
&=& 8
\textcolor{blue}{C_F T_F N_F}
\frac{(N+2)(1+N-N^2)}{N^3(N+1)^3},
\\ 
        z_{qq}^{(2)} &=&
z_{qq}^{(2),\text{NS}} +
z_{qq}^{(2),\text{PS}}
\end{eqnarray}
%---------------------------------------------------------------------------------------------
and
%---------------------------------------------------------------------------------------------
\begin{eqnarray}
R_1 =  103 N^4 + 140 N^3 + 58 N^2 + 21 N + 36.
\end{eqnarray}
%---------------------------------------------------------------------------------------------

Let us now turn now to the technical aspects of the present calculation, which has been widely automated
and the corresponding chain of programs has been described to some extent in our previous paper 
\cite{Blumlein:2021enk}. A main point concerns the observation of the current crossing relations in the 
present case, cf.~\cite{Politzer:1974fr,Blumlein:1996vs} and also Ref.~\cite{TWOLOOP}, projecting onto 
the contributions of the odd integer moments
%--------------------------------------------------------------------------------------------------------
\begin{eqnarray}
\sum_{N=0}^\infty (\Delta.k)^N \left(t^N - (-t)^N\right) \rightarrow
\left[\frac{1}{1 - \Delta.k~t} - \frac{1}{1 + \Delta.k~t}\right].
\label{eq:resu1}
\end{eqnarray}
%--------------------------------------------------------------------------------------------------------
Through this one obtains a quadratic dependence on the resummation variable $t$. An expansion in 
$t$ leads to the moments again.

Our chain of programs, from the generation of Feynman diagrams to the final result, the polarized
anomalous dimensions at three--loop order includes the packages {\tt QGRAF, FORM, Color,
EvaluateMultiSums,Crusher,SolveCoupledSystems,Guess,Sage,Sigma,HarmoncisSums} 
\cite{Nogueira:1991ex,Bierenbaum:2009mv,FORM,vanRitbergen:1998pn,EMSSP,CRUSHER,Blumlein:2019hfc,
GUESS,Blumlein:2009tj,SAGE,GSAGE,SIG1,SIG2,HARMSU,Blumlein:2009ta,Vermaseren:1998uu,Blumlein:1998if,
Remiddi:1999ew,Ablinger:2011te,Ablinger:2013cf,Ablinger:2014bra,Blumlein:2003gb,Blumlein:2009cf}. More 
details were given 
in Ref.~\cite{Blumlein:2021enk}.
We determine the contributions to the anomalous dimensions due to the different color and
multiple zeta values \cite{Blumlein:2009cf} individually. In the unfolding process also higher order terms 
$a_{ij}^{(k)}$ in the dimensional parameter $\ep$ are needed to be calculated, cf.~Ref.~\cite{TWOLOOP}.
We use the integration-by-parts relations \cite{IBP,CRUSHER} to reduce the problem to master integrals. The 
method of arbitrary high moments \cite{Blumlein:2017dxp} allows to find the needed difference equations, 
which are then solved using algorithms in difference ring theory \cite{DRING}. To calculate initial values 
for the solution we use relations given in \cite{Chetyrkin:1981qh,INIT}.

We turn now to some statistical characteristics of the present calculation.
In the polarized flavor singlet case 125 irreducible diagrams contribute for 
$\Delta A_{qq}^{(3),\rm PS}$, 1101 for 
$\Delta A_{qg}^{(3)}$, 400 for 
$\Delta A_{gq}^{(3)}$, and 1598 for 
$\Delta A_{gg}^{(3)}$. For comparison the number of diagrams for $\Delta A_{qq}^{(3),\rm NS}$ amounts to 559.
Finally, 24 diagrams contribute to the part of the forward Compton amplitude, from which $\Delta 
\gamma_{qq}^{(2), \rm s, 
NS}$ is extracted. The total number of irreducible diagrams in the polarized singlet case at three--loop 
order is larger by a factor of $\sim 22$
than in the two--loop order. The reducible diagrams 
are accounted for by wave--function renormalization \cite{Chetyrkin:2017bjc,Egorian:1978zx,
Luthe:2016xec}, decorating the OMEs at lower order in the coupling constant 
\cite{Matiounine:1998re,TWOLOOP}. 

To calculate the anomalous 
dimensions $\Delta \gamma_{\rm ij}^{(2)}$ we generated 
3000 odd moments. It turns out that the determination of the largest recurrence requires 
462 moments for $\Delta \gamma_{\rm qq}^{(2), \rm PS}$,  
989 moments for $\Delta \gamma_{\rm qg}^{(2)}$, 
1035 moments for $\Delta \gamma_{\rm gq}^{(2)}$, 
{1568} moments for $\Delta \gamma_{\rm gg}^{(2)}$.

In this way the anomalous dimensions are obtained.
The largest difference equation contributing has order {\sf o} = {16} 
and degree  {\sf d} = {304}. 
These numbers are of the order obtained in the non--singlet case in Ref.~\cite{Blumlein:2009tj}, 
where the largest difference equation contributing had order {\sf o} = 16 and degree {\sf d} = 192, 
and required 1079 moments. A moment based test--run for the polarized anomalous dimensions, like in 
the unpolarized case, has not been performed.

The overall computation time using the automated chain of codes described amounted to about {18} 
days of CPU time on {\tt Intel(R) Xeon(R) CPU E5-2643 v4} processors. Again we have only retained
the first power of the gauge parameter to have a first check on the renormalization.

The anomalous dimensions, $\Delta \gamma_{\rm ij}$, can all be expressed by harmonic sums 
\cite{Vermaseren:1998uu,Blumlein:1998if}, defined in Eq.~(34) of Ref.~\cite{Blumlein:2021enk}.
%-----------------------------------------------------------------------------------------------------
Their Mellin inversion to the splitting functions $\Delta P_{ij}(z)$ 
%-----------------------------------------------------------------------------------------------------
\begin{eqnarray}
\Delta \gamma_{ij}(N) = - \int_0^1 dz z^{N-1} \Delta P_{ij}(z) 
\label{eq:INVmel}
\end{eqnarray}
%-----------------------------------------------------------------------------------------------------
is obtained using routines of the packages {\tt HarmonicSums} and are expressed by harmonic 
polylogarithms \cite{Remiddi:1999ew}, which are defined in Eqs.~(36,37) of Ref.~\cite{Blumlein:2021enk}.

Because of the different requirements on the respective integrals, one distinguishes three contributions to 
the individual splitting functions in $z$--space
%-----------------------------------------------------------------------------------------------------
\begin{eqnarray}
\Delta P(z) = \Delta P^\delta(z)  + \Delta P^{\rm plu}(z) + \Delta P^{\rm reg}(z),
\end{eqnarray}
%-----------------------------------------------------------------------------------------------------
which are defined in Eqs.~(38,39) of Ref.~\cite{Blumlein:2021enk}.

Also here we reduce the expressions to the algebraic basis, cf.~\cite{Blumlein:2003gb}, which has 
the advantage that only the minimal set has to be calculated in numerical applications \cite{NUM}.
We will not present the splitting function in explicit form, since the expressions are rather 
lengthy. They are given in computer--readable form in an attachment to the present paper. The 
polarized singlet anomalous dimensions are given in Section~\ref{sec:4}.
%--------------------------------------------------------------------------------------------------------
\section{The polarized non--singlet  anomalous dimension \boldmath $\Delta \gamma_{\rm qq}^{(2), s, \rm NS}$} 
\label{sec:3}
%--------------------------------------------------------------------------------------------------------

\vspace*{1mm} 
\noindent 
In our previous paper \cite{Blumlein:2021enk} we had not yet calculated the non--singlet  anomalous 
dimension $\Delta \gamma_{\rm qq}^{(2), s, \rm NS}$, which emerges from three--loop order onward. It is best 
calculated using the vector--axialvector interference term in the forward Compton amplitude, 
corresponding to the associated structure function $g_5^-(x,Q^2)$, 
Ref.~\cite{Blumlein:1996vs}, corresponding to the difference in case of $W^-$ and $W^+$ 
charged current scattering. Due to its 
crossing relations it has even moments. This anomalous dimension has been calculated previously in 
Ref.~\cite{Moch:2015usa}. 

The corresponding gauge boson vertex is parameterized by
%--------------------------------------------------------------------------------------------------------
\begin{eqnarray}
i\left[ v \gamma_\mu + a \frac{i}{6} \ep_{\mu\nu_1\nu_2\nu_3} \gamma^{\nu_1} \gamma^{\nu_2} \gamma^{\nu_3}\right]
\end{eqnarray}
%--------------------------------------------------------------------------------------------------------
and we consider the current interference term $\propto a \cdot v$. The projectors for the massless external 
quark lines of momentum $p$  and boson lines corresponding to a tensor of rank two read
%--------------------------------------------------------------------------------------------------------
\begin{eqnarray}
P^q = \frac{1}{4} {\rm tr}\left[\frac{i}{6} \ep_{p\rho_1\rho_2\rho_3} 
\gamma^{\rho_1}
\gamma^{\rho_2}
\gamma^{\rho_3} ...\right];~~~~~~~P^{b}_{\mu\nu} = - \frac{g_{\mu\nu}}{D-1}.
\end{eqnarray}
%--------------------------------------------------------------------------------------------------------
The forward Compton amplitude depends on the invariants $Q^2 = -q^2$ and $p.q = Q^2/(2z) \equiv (Q^2/2) y$.
The diagrams can be represented as formal power series in $y$
%--------------------------------------------------------------------------------------------------------
\begin{eqnarray}
F(y) = \sum_{N=0}^\infty f(N) y^N,
\end{eqnarray}
%--------------------------------------------------------------------------------------------------------
cf. e.g. \cite{Blumlein:2021enk}, Sect.~2. The IBP--reduction in this case leads to three families
and 101 master integrals in total. We consider the differential equations for the individual master 
integrals $M_k(y,\ep)$, which are computed using the method described in \cite{Ablinger:2018zwz}. After insertion
of the master integrals into the amplitude and subsequent expansion in the dimensional parameter $\ep$, 
the anomalous dimension can be determined by using the command
{\tt GetMoment[F[y],y,N]} of the package {\tt HarmonicSums} from the pole term $O(1/\ep)$ of the forward Compton 
amplitude.  We obtain 
%--------------------------------------------------------------------------------------------------------
\begin{eqnarray}
\label{eq:NS2}
\Delta \gamma_{\rm NS}^{\rm (2), s} &=& - 16 \frac{1 + (-1)^N}{2} \textcolor{blue}{N_F 
\frac{d^{abc} 
d_{abc}}{N_c}}\Biggl[
        \frac{1}{N^2}
        +\frac{2 Q_1}{N^4 (1+N)^4} S_1
        +\frac{2 \big(
                2+3 N+3 N^2\big)}{N^2 (1+N)^2} 
\nonumber\\ &&
\times [S_3 - 2 S_{-3} + 4 S_{-2,1}]
        +\Biggl(
                \frac{4 \big(
                        2+4 N+4 N^2+N^3+N^4\big)}{N^3 (1+N)^3}
                +\frac{8 (-1+N) (2+N)}{N^2 (1+N)^2} S_1
        \Biggr) 
\nonumber\\ && \times
S_{-2}
\Biggr]
\end{eqnarray}
%--------------------------------------------------------------------------------------------------------
with
%--------------------------------------------------------------------------------------------------------
\begin{eqnarray}
Q_1 &=& 3 N^6+8 N^5-N^4-14 N^3-29 N^2-21 N-6.
\end{eqnarray}
%--------------------------------------------------------------------------------------------------------
The corresponding splitting function reads
%--------------------------------------------------------------------------------------------------------
\begin{eqnarray}
\label{eq:NS3}
\Delta P_{\rm NS}^{\rm (2), s} &=&  16 \textcolor{blue}{N_F \frac{d^{abc}
d_{abc}}{N_c}}\Biggl\{
        (1+x) \Biggl[
                -4 \HA_0 \HA_{-1} \big(
                        -4+5 \HA_{-1}\big)
                +8 \HA_{0,-1} \big(
                        -2+5 \HA_{-1}\big)
\nonumber\\ &&                
+18 \HA_{-1} \HA_0^2
                +\big(
                        -2 \HA_0^2
                        +32 \HA_{-1}
                \big) \HA_{0,1}
                +8 \HA_0 \HA_{0,0,1}
                -32 \HA_{0,1,-1}
                -32 \HA_{0,-1,1}
\nonumber\\ &&                 
-40 \HA_{0,-1,-1}
                -\big(
                        52 \HA_{-1}
                        +8 \HA_{0,1}
                \big) \zeta_2
        \Biggr]
        +  (1-x)\Biggl[
                24 \HA_1
                -\HA_0^2 \HA_1
                +2 \HA_0 \HA_{0,1}
\nonumber\\ &&                
 +4 \HA_0^2 \HA_{0,-1}
                +8 \HA_{0,-1}^2
                -8 \HA_0 \HA_{0,0,-1}
                -16 \HA_0 \HA_{0,-1,-1}
                -\big(
                        20 \HA_1
                        +8 \HA_{0,-1}
                \big) \zeta_2
        \Biggr]
\nonumber\\ & &
        -(1+24 x) \HA_0
        -4 x \HA_0^2
        -6 x \HA_0^3
        +\frac{1}{3} x \HA_0^4
        +2 (5+4 x) \HA_{0,1}
        -28 \HA_0 \HA_{0,-1}
\nonumber\\ &&  
       +4 (1-9 x) \HA_{0,0,1}
        +4 (5-9 x) \HA_{0,0,-1}
        +\big(
                2 (-5+4 x)
                +2 (-3+37 x) \HA_0
\nonumber\\ &&                
 -2 (3+5 x) \HA_0^2
        \big) \zeta_2
        +\big(
                2 (3+44 x)
                -16 x \HA_0
        \big) \zeta_3
        +2 (5+3 x) \zeta_2^2
\Biggr\}.
\end{eqnarray}
%--------------------------------------------------------------------------------------------------------
Here $\zeta_k$ denotes the Riemann $\zeta$--function at integer argument $k \geq 2$ and the color factor is 
normalized in the present case to $d^{abc} d_{abc}/{N_c} = 5/18$ in QCD. The leading small $z$ contribution
of $\Delta P_{\rm NS}^{\rm (2), s}$ is $\propto \ln^2(z)$. This behaviour is, however, not dominant 
in kinematic regions being accessible at present. The asymptotic behaviour reaches the complete function up 
to 10\% below $z \sim 10^{-11}$ only. One more logarithmic order allows
a description below $z \sim 10^{-3}$.   
%--------------------------------------------------------------------------------------------------------
\section{The polarized singlet anomalous dimensions} 
\label{sec:4}
%--------------------------------------------------------------------------------------------------------

\vspace*{1mm} 
\noindent 
In the following we use the minimal representations in terms of the contributing harmonic sums and harmonic 
polylogarithms by applying the algebraic relations \cite{Blumlein:2003gb} between the harmonic sums and the harmonic 
polylogarithms. The polarized singlet anomalous dimensions can be represented by the following 23 harmonic 
sums up to weight {\sf w = 5} to three--loop order
%--------------------------------------------------------------------------------------------------------
\begin{eqnarray}
&& \Biggl\{
S_{-5},S_{-4},S_{-3},S_{-2},S_1,S_2,S_3,S_4,S_5,
S_{-4,1},S_{-3,1},S_{-2,1},S_{-2,2},S_{-2,3},S_{2,-3},S_{2,1},S_{3,1},
S_{-3,1,1},
\nonumber\\ &&
S_{-2,1,1},S_{-2,2,1},S_{2,1,-2},S_{2,1,1},
S_{-2,1,1,1}\Biggr\}.
\end{eqnarray}
%--------------------------------------------------------------------------------------------------------
This number is further reduced using also the structural relations \cite{Blumlein:2009ta,Ablinger:2013jta}
to at least 15 sums. In the present case the 10 sums
%--------------------------------------------------------------------------------------------------------
\begin{eqnarray}
&& \Biggl\{
S_1, 
S_{2,1}, 
S_{-2,1}, 
S_{-3,1}, 
S_{-4,1}, 
S_{2,1,1}, 
S_{-2,1,1}.
S_{2,1,-2}, 
S_{-3,1,1}, 
S_{-2,1,1,1} 
\Biggr\}
\end{eqnarray}
%--------------------------------------------------------------------------------------------------------
suffice.

The splitting functions in $z$--space depend on the 26 harmonic polylogarithms
%--------------------------------------------------------------------------------------------------------
\begin{eqnarray}
&& \Biggl\{
\HA_{-1},\HA_0,\HA_1,
\HA_{0,-1},\HA_{0,1},
\HA_{0,-1,-1},\HA_{0,-1,1},\HA_{0,0,-1},\HA_{0,0,1},\HA_{0,1,-1},
\HA_{0,1,1},
\HA_{0,-1,-1,-1},\HA_{0,-1,-1,1},
\nonumber\\ &&
\HA_{0,-1,0,1},
\HA_{0,-1,1,-1},
\HA_{0,-1,1,1},\HA_{0,0,-1,-1},\HA_{0,0,-1,1},\HA_{0,0,0,-1},
\HA_{0,0,0,1},\HA_{0,0,1,-1},\HA_{0,0,1,1},\HA_{0,1,-1,-1},
\nonumber\\ &&
\HA_{0,1,-1,1},\HA_{0,1,1,-1},\HA_{0,1,1,1}\Biggr\}.
\end{eqnarray}
%--------------------------------------------------------------------------------------------------------
The harmonic sums are defined at the odd integers in the first place and the analytic continuation to $N \in \mathbb{C}$
is performed from there, \cite{ANCONT,Blumlein:2009ta}.

We obtain the following expressions for the polarized singlet anomalous dimensions in Mellin--$N$ space, using
the shorthand notation $S_{\vec{a}}(N) \equiv S_{\vec{a}}$ from one-- 
to three--loop order. Here we dropped
the prefactor $\tfrac{1}{2}(1 - (-1)^N)$.
%-----------------------------------------------------------------------------------------------
%\begin{eqnarray}
%\label{eq:ga1}
\begin{eqnarray}
	\Delta \gamma_{qq}^{(0)} &=& 
    \textcolor{blue}{C_F} \Biggl[
        -
        \frac{2 \big(
                2+3 N+3 N^2\big)}{N (1+N)}
        +8 S_1
\Biggr],
\\
%------------------------------------------------
\Delta \gamma_{qg}^{(0)} &=& 
-\textcolor{blue}{T_F N_F} \frac{8 (N-1)}{N (1+N)},
\\
%------------------------------------------------
\Delta \gamma_{gq}^{(0)} &=& 
-\textcolor{blue}{C_F} \frac{4 (2+N)}{N (1+N)},
\\
%------------------------------------------------
\Delta \gamma_{gg}^{(0)} &=& 
\textcolor{blue}{T_F N_F} \frac{8}{3}
+\textcolor{blue}{C_A} \Biggl[
        -\frac{2 \big(
                24+11 N+11 N^2\big)}{3 N (1+N)}
        +8 S_1
\Biggr],
\\
%------------------------------------------------
\Delta \gamma_{qq}^{(1),\rm PS} &=& 
\textcolor{blue}{C_F T_F N_F}
\frac{16 (2+N) \big(
        1+2 N+N^3\big)}{N^3 (1+N)^3},
\\
%------------------------------------------------
\Delta \gamma_{qg}^{(1)} &=& 
\textcolor{blue}{C_F T_F N_F}
\Biggl[
        -\frac{8 (-1+N) \big(
                2-N+10 N^3+5 N^4\big)}{N^3 (1+N)^3}
        +\frac{32 (N-1)}{N^2 (1+N)} S_1
\nonumber\\ &&
        -\frac{16 (N-1) }{N (1+N)} [S_1^2 - S_2]
\Biggr]
+\textcolor{blue}{C_A T_F N_F}
\Biggl[
        -\frac{16 P_{20}}{N^3 (1+N)^3}
        -\frac{64 S_1}{N (1+N)^2}
\nonumber\\ &&
        +\frac{16 (N-1) }{N (1+N)} [S_1^2 + S_2 + 2 S_{-2}]
\Biggr],
\\
%------------------------------------------------
\Delta \gamma_{gq}^{(1)} &=& 
\textcolor{blue}{C_F} \Biggl[
        \textcolor{blue}{T_F N_F}\Biggl(
                \frac{32 (2+N) (2+5 N)}{9 N (1+N)^2}
                -\frac{32 (2+N)}{3 N (1+N)} S_1
        \Biggr)
        +\textcolor{blue}{C_A} \Biggl(
                -\frac{8 P_{33}}{9 N^3 (1+N)^3}
\nonumber\\ &&                
+\frac{8 \big(
                        12+22 N+11 N^2\big) S_1}{3 N^2 (1+N)}
                -\frac{8 (2+N) S_1^2}{N (1+N)}
                +\frac{8 (2+N) S_2}{N (1+N)}
                +\frac{16 (2+N) S_{-2}
                }{N (1+N)}
        \Biggr)
\Biggr]
\nonumber\\ &&
+\textcolor{blue}{C_F^2} \Biggl[
        \frac{4 (2+N) (1+3 N) \big(
                -2-N+3 N^2+3 N^3\big)}{N^3 (1+N)^3}
        -\frac{8 (2+N) (1+3 N)}{N (1+N)^2} S_1
\nonumber\\ &&        
+\frac{8 (2+N)}{N (1+N)} [S_1^2 + S_2]
\Biggr],
\\
%------------------------------------------------
\Delta \gamma_{gg}^{(1)} &=&
\textcolor{blue}{C_F T_F N_F}
\frac{8 P_{40}}{N^3 (1+N)^3}
+\textcolor{blue}{C_A^2} \Biggl[
        -\frac{4 P_{46}}{9 N^3 (1+N)^3}
        +\Biggl(
                \frac{8 P_{17}}{9 N^2 (1+N)^2}
                -32 S_2
        \Biggr) S_1
\nonumber\\ && 
        +\frac{64}{N (1+N)} S_2
        -16 S_3
        +\Biggl(
                \frac{64}{N (1+N)}
                -32 S_1
        \Biggr) S_{-2}
        -16 S_{-3}
        +32 S_{-2,1}
\Biggr]
\nonumber\\ && 
+\textcolor{blue}{C_A T_F N_F} \Biggl[
        \frac{32 P_5}{9 N^2 (1+N)^2}
        -\frac{160}{9} S_1
\Biggr],
\\
%------------------------------------------------
\Delta \gamma_{qq}^{(2), \rm PS} &=&
\textcolor{blue}{C_F} \Biggl[
        \textcolor{blue}{T_F^2 N_F^2} \Biggl[
                -\frac{64 (2+N) P_{30}}{27 N^4 (1+N)^4}
                +\frac{64 (2+N) \big(
                        6+10 N-3 N^2+11 N^3\big)}{9 N^3 (1+N)^3} S_1
\nonumber\\ &&                
 -\frac{32 (N-1) (2+N)}{3 N^2 (1+N)^2} [S_1^2 + S_2]
        \Biggr]
        +\textcolor{blue}{C_A T_F N_F} \Biggl[
                \frac{8 P_9}{3 N^3 (1+N)^3} S_1^2
                +\frac{8 P_{10}}{3 N^3 (1+N)^3} S_2
\nonumber\\ &&
+\frac{16 P_{61}}{27 N^5 (1+N)^5}
                +\Biggl(
                        -\frac{16 P_{51}}{9 N^4 (1+N)^4}
                        +\frac{32 (N-1) (2+N)}{N^2 (1+N)^2} S_2
                \Biggr) S_1
\nonumber\\ && 
                -\frac{32 (N-1) (2+N)}{3 N^2 (1+N)^2} S_1^3
                +
                \frac{16 \big(
                        -58+23 N+23 N^2\big)}{3 N^2 (1+N)^2} S_3
                +\Biggl(
                        -\frac{32 P_1}{N^3 (1+N)^3}
\nonumber\\ &&                        
+\frac{64 (N-1) (2+N)}{N^2 (1+N)^2} S_1
                \Biggr) S_{-2}
                +\frac{32 \big(
                        -10+7 N+7 N^2\big)}{N^2 (1+N)^2} S_{-3}
                -\frac{64 (N-1) (2+N)}{N^2 (1+N)^2} S_{2,1}
\nonumber\\ &&                 
-\frac{64 \big(
                        -2+3 N+3 N^2\big)}{N^2 (1+N)^2} S_{-2,1}
                -\frac{192 (N-1) (2+N)}{N^2 (1+N)^2} \zeta_3
        \Biggr]
\Biggr]
\nonumber\\ && 
+\textcolor{blue}{C_F^2 T_F N_F} \Biggl[
        -\frac{16 (2+N) P_{54}}{N^5 (1+N)^5}
        +\Biggl(
                \frac{16 (2+N) P_{27}}{N^4 (1+N)^4}
                -\frac{32 (N-1) (2+N)}{N^2 (1+N)^2} S_2
        \Biggr) S_1
\nonumber\\ && 
        -\frac{8 (N-1) (2+N) \big(
                2+3 N+3 N^2\big)}{N^3 (1+N)^3} S_1^2
        +\frac{32 (N-1) (2+N)}{3 N^2 (1+N)^2} S_1^3
\nonumber\\ && 
-\frac{8 (2+N) \big(
                14+23 N+11 N^3\big)}{N^3 (1+N)^3} S_2
        -\frac{224 (N-1) (2+N)}{3 N^2 (1+N)^2} S_3
\nonumber\\ &&        
+\frac{64 (N-1) (2+N)}{N^2 (1+N)^2} S_{2,1}
        +\frac{192 (N-1) (2+N)}{N^2 (1+N)^2} \zeta_3
\Biggr],
\\
%-----------------------------------------------------------------------
\label{eq:qg2}
\Delta \gamma_{qg}^{(2)} &=&
\textcolor{blue}{C_F T_F^2 N_F^2} \Biggl[
                \frac{4 P_{64}}{27 N^5 (1+N)^5}
                +\Biggl(
                        -\frac{32 \big(
                                -24+4 N+47 N^2\big)}{27 N^2 (1+N)}
                        -\frac{32 (N-1)}{3 N (1+N)} S_2
                \Biggr) S_1
\nonumber\\ && 
                +\frac{32 (N-1) (3+10 N)}{9 N^2 (1+N)} S_1^2
                -
                \frac{32 (N-1)}{9 N (1+N)} S_1^3
                +\frac{32 (5 N-1)}{3 N^2 (1+N)} S_2
                +\frac{320 (N-1)}{9 N (1+N)} S_3
        \Biggr]
\nonumber\\ && 
        +\textcolor{blue}{C_A C_F T_F N_F} \Biggl[
                \frac{8 P_{29}}{3 N^3 (1+N)^3} S_2
                +\frac{P_{65}}{27 N^5 (1+N)^5 (2+N)}
                +\Biggl(
                        -\frac{384 (N-1)}{N (1+N)} S_{2,1} 
\nonumber\\ &&                         
                        +\frac{16 P_{60}}{27 N^4 (1+N)^4 (2+N)}
+\frac{16 \big(
                                75+14 N+18 N^2+N^3\big) S_2}{3 N^2 (1+N)^2}
                        +\frac{640 (N-1)}{3 N (1+N)} S_3
\nonumber\\ &&                
         -\frac{192 (N-1)}{N (1+N)} \zeta_3
                \Biggr) S_1
                +\Biggl(
                        -\frac{8 P_{25}}{9 N^3 (1+N)^3}
                        +\frac{160 (N-1)}{N (1+N)} S_2
                \Biggr) S_1^2
\nonumber\\ &&                
 +\frac{16 \big(
                        3-31 N-18 N^2+10 N^3\big)}{9 N^2 (1+N)^2} S_1^3
                +\frac{32 (N-1)}{3 N (1+N)} S_1^4
                -\frac{64 (N-1)}{N (1+N)} S_2^2
\nonumber\\ &&                
 -\frac{16 (N-1) \big(
                        240-17 N+19 N^2\big)}{9 N^2 (1+N)^2} S_3
                +\Biggl(
                        \frac{128 (N-1) \big(
                                -4-N+N^2\big)}{N^2 (1+N)^2 (2+N)} S_1
\nonumber\\ &&                         
-\frac{32 P_{24}}{N^3 (1+N)^3 (2+N)}
                        +\frac{192 (N-1)}{N (1+N)} S_1^2
                \Biggr) S_{-2}
                +\frac{96 (N-1)}{N (1+N)} S_{-2}^2
\nonumber\\ &&                
 +\frac{32 (N-1) (2+N) (-1+3 N)}{N^2 (1+N)^2} S_{-3}
                +\frac{96 (N-1) \big(
                        4+N+N^2\big) 
                }{N^2 (1+N)^2} S_{2,1}
\nonumber\\ && 
                +\frac{160 (N-1)}{N (1+N)} S_{-4}
                +\frac{64 (N-1)}{N (1+N)} S_{3,1}
                -\frac{128 (N-1)^2}{N^2 (1+N)^2} S_{-2,1}
                +\frac{64 (N-1)}{N (1+N)} S_{-2,2}
\nonumber\\ &&                
 +\frac{192 (N-1)}{N (1+N)} S_{2,1,1}
                -\frac{256 (N-1)}{N (1+N)} S_{-2,1,1}
                -\frac{192 (N-1) \big(
                        -5+3 N+3 N^2\big)}{N^2 (1+N)^2} \zeta_3
        \Biggr]
\nonumber\\ && 
+\textcolor{blue}{C_A T_F^2 N_F^2} \Biggl[
        \frac{16 P_{55}}{27 N^4 (1+N)^4}
        +\Biggl(
                \frac{64 \big(
                        23+50 N+10 N^2+19 N^3\big)}{27 N (1+N)^3}
                -\frac{32 (N-1)}{3 N (1+N)} 
\nonumber\\ &&  \times S_2
        \Biggr) 
S_1
        -\frac{64 \big(
                -2+5 N^2\big)}{9 N (1+N)^2} S_1^2
        +\frac{32 (N-1)}{9 N (1+N)} S_1^3
        -\frac{64 \big(
                -2+6 N+5 N^2\big)}{9 N (1+N)^2} S_2
\nonumber\\ &&         
+\frac{64 (N-1)}{9 N (1+N)} S_3
        -\frac{128 (-2+5 N)}{9 N (1+N)} S_{-2}
        +\frac{128 (N-1) }{3 N (1+N)} [S_{-3} + S_{2,1}]
\Biggr] 
\nonumber\\ &&
+\textcolor{blue}{C_A^2 N_F T_F} \Biggl[
        \frac{16  P_{34}}{9 N^3 (1+N)^3} S_2
        -\frac{8 P_{70}}{27 N^5 (1+N)^5 (2+N)}
        +\Biggl(
                -\frac{8 P_{56}}{27 N^4 (1+N)^4}
\nonumber\\ &&                 
+\frac{8 \big(
                        -72+181 N-48 N^2+11 N^3\big) }{3 N^2 (1+N)^2} S_2
                -\frac{704 (N-1)}{3 N (1+N)} S_3
                +\frac{128 (N-1)}{N (1+N)} S_{2,1}
\nonumber\\ &&                 
+
                \frac{512 (N-1)}{N (1+N)} S_{-2,1}
                +\frac{192 (N-1)}{N (1+N)} \zeta_3
        \Biggr) S_1
        +\Biggl(
                \frac{16 P_{31}}{9 N^3 (1+N)^3}
                -\frac{160 (N-1)}{N (1+N)} S_2
        \Biggr) S_1^2
\nonumber\\ &&         
-\frac{8 \big(
                -24-59 N+11 N^3\big)}{9 N^2 (1+N)^2} S_1^3
        -\frac{16 (N-1)}{3 N (1+N)} S_1^4
        -\frac{16 (N-1)}{N (1+N)} S_2^2
\nonumber\\ &&         
-\frac{16 \big(
                345-428 N+11 N^3\big)}{9 N^2 (1+N)^2} S_3
        -\frac{32 (N-1)}{N (1+N)} S_4
        +\Biggl(
                \frac{32 P_{49}}{9 N^3 (1+N)^3 (2+N)}
\nonumber\\ &&             
    -\frac{64 (-5+N) (-1+2 N) }{N^2 (1+N)^2} S_1
                -\frac{192 (N-1)}{N (1+N)} S_1^2
                -\frac{128 (N-1)}{N (1+N)} S_2
        \Biggr) S_{-2}
\nonumber\\ &&         
        -\frac{96 (N-1)}{N (1+N)} S_{-2}^2
+\Biggl(
                -\frac{32 \big(
                        69-92 N+11 N^3\big)}{3 N^2 (1+N)^2}
                -\frac{512 (N-1)}{N (1+N)} S_1
        \Biggr) S_{-3}
\nonumber\\ && 
        -\frac{352 (N-1)}{N (1+N)} S_{-4}  
        -\frac{32 (N-1) \big(
                24+11 N+11 N^2\big)}{3 N^2 (1+N)^2} S_{2,1}
        -\frac{128 (N-1)}{N (1+N)} S_{3,1}
\nonumber\\ &&         
-\frac{64 (-7+11 N)}{N^2 (1+N)^2} S_{-2,1}
        +\frac{448 (-1+N) }{N (1+N)} S_{-2,2}
        +\frac{512 (-1+N) }{N (1+N)} S_{-3,1}
\nonumber\\ &&         
-\frac{768 (-1+N) }{N (1+N)} S_{-2,1,1}
        +\frac{96 (-1+N) \big(
                -8+3 N+3 N^2\big)}{N^2 (1+N)^2} \zeta_3
\Biggr]
\nonumber\\ && 
+\textcolor{blue}{C_F^2 T_F N_F} \Biggl[
        -
        \frac{8 P_{26}}{N^3 (1+N)^3} S_1^2
        +\frac{8 P_{28}}{N^3 (1+N)^3} S_2
+\frac{P_{63}}{N^4 (1+N)^5 (2+N)}
\nonumber\\ &&         
        +\Biggl(
                -\frac{8 P_{53}}{N^4 (1+N)^4}
                -\frac{8 \big(
                        -6+7 N+28 N^2+3 N^3\big)}{N^2 (1+N)^2} S_2
                -\frac{704 (N-1)}{3 N (1+N)} S_3
\nonumber\\ &&                
 +\frac{256 (N-1)}{N (1+N)} S_{2,1}
        \Biggr) S_1
        -\frac{8 (N-1) \big(
                -10-9 N+3 N^2\big)}{3 N^2 (1+N)^2} S_1^3
        -\frac{16 (N-1)}{3 N (1+N)} S_1^4
\nonumber\\ &&         
-\frac{48 (N-1)}{N (1+N)} S_2^2
        -\frac{16 (N-1) \big(
                -22+27 N+3 N^2\big)}{3 N^2 (1+N)^2} S_3
        -\frac{160 (N-1)}{N (1+N)} S_4
\nonumber\\ &&         
+\Biggl(
                \frac{64 P_{21}}{N^2 (1+N)^3 (2+N)}
                -\frac{256 (N-1)}{N (1+N)^2} S_1
                -\frac{128 (N-1)}{N (1+N)} S_2
        \Biggr) S_{-2}
        -\frac{64 (N-1)}{N (1+N)} S_{-2}^2
\nonumber\\ &&         
+\Biggl(
                -\frac{128 (N-1)^2}{N^2 (1+N)^2}
                -\frac{256 (N-1) }{N (1+N)} S_1
        \Biggr) S_{-3}
        -\frac{320 (N-1)}{N (1+N)} S_{-4}
        -\frac{128 (N-1)}{N^2 (1+N)^2} S_{2,1}
\nonumber\\ &&         
+\frac{64 (N-1)}{N (1+N)} S_{3,1}
        +\frac{256 (N-1)}{N (1+N)^2} S_{-2,1}
        +\frac{128 (N-1)}{N (1+N)} S_{-2,2}
        +\frac{256 (N-1)}{N (1+N)} S_{-3,1}
\nonumber\\ &&         
-\frac{192 (N-1)}{N (1+N)} S_{2,1,1}
        +\frac{96 (N-1) \big(
                -2+3 N+3 N^2\big)}{N^2 (1+N)^2} \zeta_3
\Biggr],
\\
%--------------------------------------------------------------------------------------------------
\Delta \gamma_{gq}^{(2)} &=&
\textcolor{blue}{C_F^2} \Biggl[
        \textcolor{blue}{C_A} \Biggl[
                \frac{4 P_{39}}{9 N^3 (1+N)^3} S_2
                +\frac{P_{73}}{54 (N-1) N^5 (1+N)^5}
                +\Biggl(
                        -\frac{4 P_{58}}{27 N^4 (1+N)^4}
\nonumber\\ &&                         
-\frac{8 \big(
                                30+203 N+177 N^2+49 N^3\big)}{3 N^2 (1+N)^2} S_2
                        -\frac{640 (2+N)}{3 N (1+N)} S_3
                        +\frac{64 (2+N)}{N (1+N)} S_{2,1}
\nonumber\\ &&                
         +\frac{128 (2+N)}{N (1+N)} S_{-2,1}
                        +\frac{288 (2+N)}{N (1+N)} \zeta_3
                \Biggr) S_1
                +\Biggl(
                        \frac{4 P_{36}}{9 N^3 (1+N)^3}
                        -\frac{16 (2+N)}{N (1+N)} S_2
                \Biggr) S_1^2
\nonumber\\ &&               
 -\frac{8 \big(
                        6+85 N+132 N^2+50 N^3\big)}{9 N^2 (1+N)^2} S_1^3
                +\frac{8 (2+N) \big(
                        102+65 N+29 N^2\big)}{9 N^2 (1+N)^2} S_3
\nonumber\\ && 
                +\frac{16 (2+N)}{3 N (1+N)} S_1^4
                -\frac{32 (2+N)}{N (1+N)} S_4
                +\Biggl(
                        -\frac{16 P_{44}}{(N-1) N^3 (1+N)^3}
                        +\frac{64 (7+3 N)}{N (1+N)} S_1
\nonumber\\ &&                
         -\frac{96 (2+N)}{N (1+N)} S_1^2
                \Biggr) S_{-2}
                +\Biggl(
                        \frac{32 (2+N) (4+3 N)}{N (1+N)^2}
                        -\frac{64 (2+N)}{N (1+N)} S_1
                 \Biggr) S_{-3}
\nonumber\\ && 
                +\frac{80 (2+N)}{N (1+N)} [S_{-2}^2 + S_{-4}]
                +\frac{16 (2+N) \big(
                        -6+11 N+11 N^2\big)}{3 N^2 (1+N)^2} S_{2,1}
                +\frac{224 (2+N)}{N (1+N)} S_{3,1}
\nonumber\\ &&                
 -\frac{32 (2+N) (5+3 N)}{N (1+N)^2} S_{-2,1}
                +\frac{32 (2+N)}{N (1+N)} S_{-2,2}
                -
                \frac{96 (2+N)}{N (1+N)} S_{2,1,1}
\nonumber\\ &&               
 -\frac{128 (2+N)}{N (1+N)} S_{-2,1,1}
                -\frac{432 (2+N) \zeta_3}{N (1+N)}
        \Biggr]
        +\textcolor{blue}{T_F N_F} \Biggl[
                \frac{2 P_{71}}{27 (N-1) N^5 (1+N)^5}
\nonumber\\ &&                 
+\Biggl(
                        \frac{32 (2+N) P_{14}}{27 N^3 (1+N)^3}
                        +\frac{208 (2+N)}{3 N (1+N)} S_2
                \Biggr) S_1
                -\frac{16 (2+N) \big(
                        -3+16 N+37 N^2\big)}{9 N^2 (1+N)^2} S_1^2
\nonumber\\ &&                
 +\frac{80 (2+N)}{9 N (1+N)} S_1^3
                -\frac{16 (2+N) \big(
                        9+46 N+67 N^2\big)}{9 N^2 (1+N)^2} S_2
                +\frac{256 (2+N)}{9 N (1+N)} S_3
\nonumber\\ &&               
 +\frac{256}{(N-1) N^2 (1+N)^2} S_{-2}
                -\frac{64 (2+N)}{3 N (1+N)} S_{2,1}
                -\frac{128 (2+N)}{N (1+N)} \zeta_3
        \Biggr]
\Biggr]
\nonumber\\ && 
+\textcolor{blue}{C_F} \Biggl[
        \textcolor{blue}{T_F^2 N_F^2} \Biggl[
                \frac{64 (2+N) \big(
                        3+7 N+N^2\big)}{9 N (1+N)^3}
                +\frac{64 (2+N) (2+5 N)}{9 N (1+N)^2} S_1
\nonumber\\ &&               
  -\frac{32 (2+N)}{3 N (1+N)} [S_1^2 + S_2]
        \Biggr]
        +\textcolor{blue}{C_A T_F N_F} \Biggl[
                \frac{8 P_{57}}{27 (N-1) N^3 (1+N)^4}
\nonumber\\ &&                
 +\Biggl(
                        -\frac{16 P_{37}}{27 N^3 (1+N)^3}
                        +\frac{80 (2+N) S_2}{3 N (1+N)}
                \Biggr) S_1
                +\frac{16 \big(
                        18+116 N+129 N^2+43 N^3\big)}{9 N^2 (1+N)^2} S_1^2
\nonumber\\ &&                
 -\frac{80 (2+N)}{9 N (1+N)} S_1^3
                +
                \frac{16 \big(
                        -2+16 N+9 N^2+N^3\big)}{3 N^2 (1+N)^2} S_2
                +\frac{512 (2+N)}{9 N (1+N)} S_3
\nonumber\\ &&                
 +\Biggl(
                        -\frac{64 P_7}{3 (-1+N) N^2 (1+N)^2}
                        +\frac{256 (2+N) S_1}{3 N (1+N)}
                \Biggr) S_{-2}
                +\frac{128 (2+N) }{3 N (1+N)} [S_{-3} - S_{-2,1}]
\nonumber\\ &&                
 +\frac{128 (2+N)}{N (1+N)} \zeta_3
        \Biggr]
        +\textcolor{blue}{C_A^2} \Biggl[
                \frac{2 P_{35}}{3 N^3 (1+N)^3} S_2
                -\frac{4 P_{72}}{27 (N-1) N^5 (1+N)^5}
\nonumber\\ && 
                +\Biggl(
                        \frac{4 P_{62}}{27 (N-1) N^4 (1+N)^4}
         -\frac{4 \big(
                                120+158 N+141 N^2+55 N^3\big)}{3 N^2 (1+N)^2} S_2
                        +\frac{128 (2+N)}{3 N (1+N)} S_3 
\nonumber\\ &&                
         +\frac{128 (2+N)}{N (1+N)} S_{-2,1}
                        -\frac{96 (2+N)}{N (1+N)} \zeta_3
                \Biggr) S_1
                +\Biggl(
                        -\frac{2 P_{38}}{9 N^3 (1+N)^3}
                        +\frac{48 (2+N)}{N (1+N)} S_2
                \Biggr) S_1^2
\nonumber\\ &&                
 +\frac{4 \big(
                        24+158 N+165 N^2+55 N^3\big)}{9 N^2 (1+N)^2} S_1^3
                -\frac{8 (2+N)}{3 N (1+N)} S_1^4
                -\frac{40 (2+N)}{N (1+N)} S_2^2
\nonumber\\ &&                
 -\frac{8 \big(
                        -186+295 N+528 N^2+176 N^3\big)}{9 N^2 (1+N)^2} S_3
                -\frac{16 (2+N)}{N (1+N)} S_4
\nonumber\\ &&                
 +\Biggl(
                        -\frac{32 P_{12}}{3 (N-1) N^2 (1+N)^2} S_1
                        +\frac{16 P_{47}}{3 (N-1) N^3 (1+N)^3}
                        +\frac{96 (2+N)}{N (1+N)} S_1^2
\nonumber\\ &&                
         -
                        \frac{64 (2+N)}{N (1+N)} S_2
                \Biggr) S_{-2}
                -\frac{16 (2+N)}{N (1+N)} S_{-2}^2
                +\Biggl(
                        -\frac{16 \big(
                                -126-13 N+66 N^2+22 N^3\big)}{3 N^2 (1+N)^2}
\nonumber\\ &&                
         -\frac{192 (2+N)}{N (1+N)} S_1
                \Biggr) S_{-3}
                -\frac{176 (2+N)}{N (1+N)} S_{-4}
                +\frac{32 \big(
                        -30+13 N+33 N^2+11 N^3\big)}{3 N^2 (1+N)^2} S_{-2,1}
\nonumber\\ && 
                -\frac{64 (2+N)}{N (1+N)} S_{3,1}
                +\frac{224 (2+N)}{N (1+N)} S_{-2,2}
                +\frac{256 (2+N)}{N (1+N)} S_{-3,1}
 -\frac{384 (2+N)}{N (1+N)} S_{-2,1,1}
\nonumber\\ && 
                +\frac{144 (2+N)}{N (1+N)} \zeta_3
        \Biggr]
\Biggr]
+\textcolor{blue}{C_F^3} \Biggl[
        -\frac{2 (2+N) P_{19}}{N^3 (1+N)^3} S_2
        +\frac{P_{66}}{2 (N-1) N^5 (1+N)^5}
\nonumber\\ &&         
+\Biggl(
                -\frac{4 (2+N) P_{43}}{N^4 (1+N)^4}
                +\frac{4 (2+N) \big(
                        -2+19 N+39 N^2\big)}{N^2 (1+N)^2} S_2
                +\frac{128 (2+N)}{3 N (1+N)} S_3
\nonumber\\ &&                
 -\frac{64 (2+N)}{N (1+N)} S_{2,1}
                -\frac{192 (2+N)}{N (1+N)} \zeta_3
        \Biggr) S_1
        +\Biggl(
                \frac{2 (2+N) P_3}{N^3 (1+N)^3}
                -\frac{32 (2+N)}{N (1+N)} S_2
        \Biggr) S_1^2 
\nonumber\\ && 
+\frac{4 (2+N) \big(
                -2+3 N+15 N^2\big)}{3 N^2 (1+N)^2} S_1^3
        -\frac{8 (2+N)}{3 N (1+N)} S_1^4
        -\frac{24 (2+N)}{N (1+N)} S_2^2
\nonumber\\ &&         
+\frac{64 (2+N) \big(
                -1+3 N^2\big)}{3 N^2 (1+N)^2} S_3
        -
        \frac{48 (2+N)}{N (1+N)} S_4
        +\Biggl(
                \frac{32 P_{22}}{(N-1) N^2 (1+N)^3}
\nonumber\\ &&                
 -\frac{128 (2+N)}{N (1+N)^2} S_1
                -\frac{64 (2+N)}{N (1+N)} S_2
        \Biggr) S_{-2}
        -\frac{16 (2+N) \big(
                -2+3 N+3 N^2\big)}{N^2 (1+N)^2} S_{2,1}
\nonumber\\ &&         
        -\frac{96 (2+N)}{N (1+N)} S_{-2}^2 
+\Biggl(
                -\frac{64 (N-1) (2+N)}{N^2 (1+N)^2}
                -\frac{128 (2+N)}{N (1+N)} S_1
        \Biggr) S_{-3}
\nonumber\\ && 
        -\frac{160 (2+N)}{N (1+N)} [S_{-4} + S_{3,1}]
        +\frac{128 (2+N)}{N (1+N)^2} [S_{-2,1} + (1+N) S_{-3,1}]
\nonumber\\ &&       
        +\frac{64 (2+N)}{N (1+N)} S_{-2,2}
 +\frac{96 (2+N)}{N (1+N)} S_{2,1,1}
        +\frac{288 (2+N) \zeta_3}{N (1+N)}
\Biggr],
\\
%---------------------------------------------------------------------------------
\label{eq:gg2}
\Delta \gamma_{gg}^{(2)} &=&
\textcolor{blue}{C_A T_F^2 N_F^2} \Biggl[
        -\frac{16 P_8}{27 N^2 (1+N)^2} S_1
        -\frac{4 P_{48}}{27 N^3 (1+N)^3}
\Biggr]
+\textcolor{blue}{C_F} \Biggl[
        \textcolor{blue}{T_F^2 N_F^2} \Biggl[
                -\frac{8 P_{59}}{27 N^4 (1+N)^4}
\nonumber\\ &&                
+\frac{64 (N-1) (2+N) \big(
                        -6-8 N+N^2\big)}{9 N^3 (1+N)^3} S_1
                +\frac{32 (N-1) (2+N)}{3 N^2 (1+N)^2} S_1^2
\nonumber\\ && 
                -\frac{32 (N-1) (2+N)}{N^2 (1+N)^2} S_2
        \Biggr]
        +\textcolor{blue}{C_A T_F N_F} \Biggl[
                \frac{8 P_6}{N^3 (1+N)^3} S_2
                -\frac{8 P_9}{3 N^3 (1+N)^3} S_1^2
\nonumber\\ &&  
                +
                \frac{2 P_{77}}{27 (N-1) N^5 (1+N)^5 (2+N)}
                +\Biggl(
                        -\frac{8 P_{67}}{9 (-1+N) N^4 (1+N)^4 (2+N)}
\nonumber\\ &&                        
-\frac{32 (N-1) (2+N)}{N^2 (1+N)^2} S_2
                        +128 \zeta_3
                \Biggr) S_1
                +\frac{32 (N-1) (2+N)}{3 N^2 (1+N)^2} S_1^3
                -\frac{32 \big(
                        34+N+N^2\big)}{3 N^2 (1+N)^2} 
\nonumber\\ && \times
S_3
                +\Biggl(
                        \frac{128P_2}{(N-1) N^2 (1+N)^2 (2+N)} S_1
                        -\frac{32 P_{23}}{(N-1) N^2 (1+N)^3 (2+N)}
                \Biggr) S_{-2}
\nonumber\\ &&                
 -\frac{192 \big(
                        4-N-N^2\big)}{N^2 (1+N)^2} S_{-3}
                +\frac{64 (N-1) (2+N)}{N^2 (1+N)^2} S_{2,1}
                -\frac{128 \big(
                        -8+N+N^2\big)}{N^2 (1+N)^2} S_{-2,1}
\nonumber\\ &&               
  -\frac{64 (-3+N) (4+N)}{N^2 (1+N)^2} \zeta_3
        \Biggr]
\Biggr]
+\textcolor{blue}{C_A^3} \Biggl[
        \frac{64 P_{16}}{9 N^2 (1+N)^2} S_{-2,1}
        -\frac{32 P_{18}}{9 N^2 (1+N)^2} S_3
\nonumber\\ &&         
+\frac{P_{74}}{27 (N-1) N^5 (1+N)^5 (2+N)}
        +\Biggl(
                 \frac{4 P_{69}}{9 (N-1) N^4 (1+N)^4 (2+N)}
\nonumber\\ && 
                -\frac{64 P_{17}}{9 N^2 (1+N)^2} S_2
                +128 S_2^2
                +\frac{16 \big(
                        -96+11 N+11 N^2\big)}{3 N (1+N)} S_3
                +192 S_4
\nonumber\\ &&                
 +\frac{1024}{N (1+N)} S_{-2,1}
                -640 S_{-2,2}
                -768 S_{-3,1}
                +1024 S_{-2,1,1}
        \Biggr) S_1
\nonumber\\ &&         
+\Biggl(
                -\frac{256 \big(
                        1+3 N+3 N^2\big)}{N^3 (1+N)^3}
                +128 S_3
                -256 S_{-2,1}
        \Biggr) S_1^2
        +\Biggl(
                -
                \frac{16 P_{41}}{9 N^3 (1+N)^3}
\nonumber\\ &&                
 +64 S_3
                +640 S_{-2,1}
        \Biggr) S_2
        -\frac{256}{N (1+N)} S_2^2
        -\frac{384}{N (1+N)} S_4
        +64 S_5
\nonumber\\ &&         
+\Biggl(
                \frac{32 P_{52}}{9 (N-1) N^3 (1+N)^3 (2+N)}
                +\big(
                        -\frac{64 P_{32}}{9 (-1+N) N (1+N)^2 (2+N)}
                        +256 S_2
                \Biggr) 
\nonumber\\ &&  \times
S_1
                -\frac{512}{N (1+N)} S_2
                +128 S_3
                -768 S_{2,1}
        \Biggr) S_{-2}
        +\Biggl(
                -\frac{16 \big(
                        24+11 N+11 N^2\big)}{3 N (1+N)}
\nonumber\\ &&                 
+64 S_1
        \Biggr) S_{-2}^2
        +\Biggl(
                -\frac{32 P_{15}}{9 N^2 (1+N)^2}
                -\frac{1536}{N (1+N)} S_1
                +384 S_1^2
                -320 S_2
        \Biggr) S_{-3}
\nonumber\\ &&         
+\Biggl(
                -\frac{1024}{N (1+N)}
                +512 S_1
        \Biggr) S_{-4}
        -192 S_{-5}
        -384 S_{2,-3}
        +\frac{1280}{N (1+N)} S_{-2,2}
\nonumber\\ &&         
+384 S_{-2,3}
        +\frac{1536}{N (1+N)} S_{-3,1}
        -384 S_{-4,1}
        +768 S_{2,1,-2}
        -\frac{2048}{N (1+N)} S_{-2,1,1}
\nonumber\\ &&         
+768 [S_{-2,2,1} + S_{-3,1,1}]
        -1536 S_{-2,1,1,1}
\Biggr]
\nonumber\\ &&
+\textcolor{blue}{C_F^2 T_F N_F} \Biggl[
        -\frac{4 P_{75}}{(N-1) N^5 (1+N)^5 (2+N)}
        +\Biggl(
                \frac{32 (N-1) (2+N) S_2}{N^2 (1+N)^2}
\nonumber\\ &&                 
-\frac{16 P_{42}}{N^4 (1+N)^4}
        \Biggr) S_1
        +\frac{8 (N-1) (2+N) \big(
                2+3 N+3 N^2\big)}{N^3 (1+N)^3} S_1^2
        -\frac{32 (N-1) (2+N)}{3 N^2 (1+N)^2} 
\nonumber\\ && \times S_1^3         
-\frac{8 (2+N) \big(
                2-11 N-16 N^2+9 N^3\big)}{N^3 (1+N)^3} S_2
        +\frac{32 \big(
                10+7 N+7 N^2\big)}{3 N^2 (1+N)^2} S_3
\nonumber\\ &&         
+\Biggl(
                -\frac{64 \big(
                        10+N+N^2\big)}{(N-1) N (1+N) (2+N)}
                +\frac{512}{N^2 (1+N)^2} S_1
        \Biggr) S_{-2}
        +\frac{256}{N^2 (1+N)^2} S_{-3}
\nonumber\\ &&         
-\frac{64 (N-1) (2+N)}{N^2 (1+N)^2} S_{2,1}
        -\frac{512}{N^2 (1+N)^2} S_{-2,1}
        +\frac{192 \big(
                -2-N-N^2\big)}{N^2 (1+N)^2} \zeta_3
\Biggr]
\nonumber\\ && 
+\textcolor{blue}{C_A^2 T_F N_F} \Biggl[
        \frac{32 P_4}{9 N^2 (1+N)^2} S_2
        +\frac{32 P_{11}}{9 N^2 (1+N)^2} S_{-3}
        -\frac{64 P_{11}}{9 N^2 (1+N)^2} S_{-2,1}
\nonumber\\ &&         
+\frac{16 P_{13}}{9 N^2 (1+N)^2} S_3
        +\frac{2 P_{76}}{27 (N-1) N^5 (1+N)^5 (2+N)}
        +\Biggl(
                \frac{1280}{9} S_2
                -\frac{64}{3} S_3
\nonumber\\ && 
                -\frac{8 P_{68}}{27 (-1+N) N^4 (1+N)^4 (2+N)}
                -128 \zeta_3
        \Biggr) S_1
        +\frac{64}{3} S_{-2}^2
\nonumber\\ &&         +\Biggl(
                \frac{64 P_{45}}{9 (N-1) N^2 (1+N)^2 (2+N)} 
S_1
                -\frac{32 P_{50}}{9 (N-1) N^3 (1+N)^3 (2+N)}
        \Biggr) S_{-2}
\nonumber\\ &&
        +\frac{128 \big(
                -3+2 N+2 N^2\big)}{N^2 (1+N)^2} \zeta_3
\Biggr],
\end{eqnarray}
%-----------------------------------------------------------------------------------------------------

with the polynomials
%\label{eq:ga2}
%-----------------------------------------------------------------------------------------------------
\begin{eqnarray}
P_1 &=& N^4-2 N^3-4 N^2+15 N+2,
\\
P_2 &=&N^4+2 N^3-5 N^2-6 N+16,
\\
P_3 &=& N^4+44 N^3+45 N^2+38 N+12,
\\
P_4 &=& 3 N^4+6 N^3-89 N^2-92 N+12,
\\
P_5 &=& 3 N^4+6 N^3+16 N^2+13 N-3,
\\
P_6 &=& 3 N^4+18 N^3+17 N^2-46 N-28,
\\
P_7 &=& 5 N^4+9 N^3-4 N^2-4 N+6,
\\
P_8 &=& 8 N^4+16 N^3-19 N^2-27 N+48,
\\
P_9 &=& 11 N^4+22 N^3+13 N^2+2 N-12,
\\
P_{10} &=& 11 N^4+34 N^3+N^2-70 N-12,
\\
P_{11} &=& 20 N^4+40 N^3+11 N^2-9 N+54,
\\
P_{12} &=& 22 N^4+50 N^3+5 N^2-47 N+6,
\\
P_{13} &=& 40 N^4+80 N^3+73 N^2+33 N+54,
\\
P_{14} &=& 62 N^4-17 N^3-76 N^2-69 N-18,
\\
P_{15} &=& 67 N^4+134 N^3+49 N^2+54 N-360,
\\
P_{16} &=& 67 N^4+134 N^3+49 N^2+54 N-72,
\\
P_{17} &=& 67 N^4+134 N^3+67 N^2+144 N+72,
\\
P_{18} &=& 67 N^4+134 N^3+109 N^2+114 N-126,
\\
P_{19} &=& 95 N^4+148 N^3+35 N^2-38 N-12,
\\
P_{20} &=& N^5+N^4-4 N^3+3 N^2-7 N-2,
\\
P_{21} &=& 2 N^5+6 N^4-N^3-8 N^2+15 N+10,
\\
P_{22} &=& 2 N^5+6 N^4+5 N^3+4 N^2+9 N-2,
\\
P_{23} &=& 3 N^5+5 N^4-33 N^3-45 N^2+6 N-16,
\\
P_{24} &=& 3 N^5+14 N^4+21 N^3+20 N^2+10 N+4,
\\
P_{25} &=& 8 N^5-31 N^4+205 N^3-59 N^2-447 N-108,
\\
P_{26} &=& 14 N^5+15 N^4-19 N^3-13 N^2-25 N-20,
\\
P_{27} &=& 21 N^5+9 N^4+13 N^3-13 N^2-22 N-12,
\\
P_{28} &=& 26 N^5+41 N^4-21 N^3+21 N^2+9 N-12,
\\
P_{29} &=& 36 N^5-55 N^4-243 N^3-75 N^2-163 N-108,
\\
P_{30} &=& 58 N^5+7 N^4+59 N^3+50 N^2+3 N-9,
\\
P_{31} &=& 67 N^5+49 N^4-52 N^3+164 N^2-90 N-72,
\\
P_{32} &=& 67 N^5+201 N^4+67 N^3-57 N^2+109 N-189,
\\ 
P_{33} &=& 76 N^5+271 N^4+254 N^3+41 N^2+72 N+36,
\\
P_{34} &=& 85 N^5+151 N^4-40 N^3+164 N^2-306 N-72,
\\
P_{35} &=& 171 N^5+552 N^4+343 N^3+246 N^2+1052 N+480,
\\
P_{36} &=& 305 N^5+989 N^4+907 N^3-5 N^2+60 N+36,
\\
P_{37} &=& 418 N^5+1525 N^4+1763 N^3+650 N^2+444 N+144,
\\
P_{38} &=& 631 N^5+2524 N^4+3743 N^3+3398 N^2+2844 N+864,
\\
P_{39} &=& 725 N^5+2831 N^4+3481 N^3+1699 N^2+1044 N+324,
\\
P_{40} &=& N^6+3 N^5+5 N^4+N^3-8 N^2+2 N+4,
\\
P_{41} &=& 3 N^6+9 N^5-584 N^4-1183 N^3-275 N^2-834 N-432,
\\
P_{42} &=& 5 N^6+23 N^5+11 N^4-39 N^3-20 N^2+16 N+8,
\\
P_{43} &=& 9 N^6+56 N^5+87 N^4+54 N^3-52 N^2-50 N-12,
\\
P_{44} &=& 14 N^6+39 N^5+14 N^4-19 N^3+6 N^2+14 N+4,
\\
P_{45} &=& 20 N^6+60 N^5+11 N^4-78 N^3-13 N^2+36 N-108,
\\
P_{46} &=& 48 N^6+144 N^5+469 N^4+698 N^3+7 N^2+258 N+144,
\\
P_{47} &=& 76 N^6+217 N^5+25 N^4-181 N^3+106 N^2-21 N-6,
\\
P_{48} &=& 87 N^6+261 N^5+249 N^4+63 N^3-76 N^2-64 N-96,
\\
P_{49} &=& 94 N^6+315 N^5+145 N^4-63 N^3+148 N^2-441 N-18,
\\
P_{50} &=& 95 N^6+285 N^5+92 N^4-291 N^3-97 N^2+96 N-36,
\\
P_{51} &=& 160 N^6+438 N^5+364 N^4+330 N^3+529 N^2+321 N+18,
\\
P_{52} &=& 325 N^6+975 N^5+85 N^4-879 N^3+598 N^2-240 N-72,
\\
P_{53} &=& 17 N^7+9 N^6-95 N^5-19 N^4+76 N^3+22 N^2+26 N+28,
\\
P_{54} &=& 24 N^7+33 N^6+13 N^5-28 N^4-31 N^3-33 N^2-26 N-8,
\\
P_{55} &=& 165 N^7+330 N^6-491 N^5-365 N^4-136 N^3-445 N^2-18 N+144,
\\
P_{56} &=& 475 N^7+833 N^6+1527 N^5+2905 N^4-1342 N^3+5562 N^2+3834 N+486,
\\
P_{57} &=& 537 N^7+1200 N^6-1013 N^5-2085 N^4+1720 N^3-855 N^2-2468 N-492,
\\
P_{58} &=& 1199 N^7+3523 N^6+681 N^5-5953 N^4-4214 N^3-4800 N^2-3168 N-864,
\\
P_{59} &=& 33 N^8+132 N^7+70 N^6-612 N^5-839 N^4+480 N^3+712 N^2+408 N+144,
\\
P_{60} &=& 476 N^8+2297 N^7+2018 N^6-4915 N^5-7324 N^4+242 N^3-1218 N^2
\nonumber\\ &&
-2700 N-864,
\\
P_{61} &=& 914 N^8+3005 N^7+3368 N^6+4349 N^5+5183 N^4+548 N^3+1101 N^2
\nonumber\\ &&
+936 N+324,
\\
P_{62} &=& 2078 N^8+8225 N^7+10475 N^6-1921 N^5-10729 N^4-2560 N^3-5658 N^2-7578 N
\nonumber\\ &&
-2700,
\\
P_{63} &=& -5 N^9-25 N^8+228 N^7+926 N^6-201 N^5-2377 N^4+626 N^3+2788 N^2+2168 N
\nonumber\\ &&
+480,
\\
P_{64} &=& 99 N^9+297 N^8-982 N^7-662 N^6+1035 N^5-3079 N^4+3448 N^3+2868 N^2
\nonumber\\ &&
-2448 N-1728,
\\
P_{65} &=& -1251 N^{10}-6255 N^9-10972 N^8-17422 N^7-1423 N^6+111905 N^5+149894 N^4
\nonumber\\ &&
-2116 N^3-37752 N^2+12384 N+6912,
\\
P_{66} &=& -115 N^{10}-383 N^9+356 N^8+2762 N^7+3001 N^6-471 N^5-882 N^4+1068 N^3
\nonumber\\ &&
+1352 N^2-288 N-256,
\\
P_{67} &=& 165 N^{10}+825 N^9+1102 N^8-578 N^7-1939 N^6-239 N^5+1184 N^4+448 N^3
\nonumber\\ &&
-2456 N^2-2256 N-864,
\\
P_{68} &=& 418 N^{10}+2090 N^9+3857 N^8+5096 N^7+6254 N^6-808 N^5-10295 N^4-5622 N^3
\nonumber\\ &&
+2898 N^2+2376 N+648,
\\
P_{69} &=& 735 N^{10}+3675 N^9+6060 N^8+6934 N^7+11743 N^6-41 N^5-18290 N^4-920 N^3
\nonumber\\ &&
-8168 N^2-10656 N-3744,
\\
P_{70} &=& 741 N^{10}+3705 N^9+2650 N^8-8780 N^7-12083 N^6-13127 N^5-15536 N^4+3586 N^3
\nonumber\\ &&
-16128 N^2-11916 N-3240,
\\
P_{71} &=& 1065 N^{10}+6693 N^9+14084 N^8+10058 N^7-3475 N^6-11707 N^5+446 N^4+17132 N^3
\nonumber\\ &&
+3432 N^2-6624 N-3456,
\\
P_{72} &=& 3321 N^{10}+13584 N^9+9571 N^8-17159 N^7-7838 N^6+5281 N^5-20690 N^4-842 N^3
\nonumber\\ &&
-5208 N^2-7884 N-3240,
\\
P_{73} &=& 18579 N^{10}+68775 N^9+3212 N^8-235282 N^7-220465 N^6+54263 N^5+91994 N^4
\nonumber\\ &&
-48748 N^3-24648 N^2+29664 N+13824,
\\
P_{74} &=& -2133 N^{12}-12798 N^{11}-54337 N^{10}-153794 N^9-137083 N^8+105398 N^7
\nonumber\\ &&
+31109 N^6-29734 N^5+318684 N^4-18512 N^3+82224 N^2+126720 N+48384,
\\
P_{75} &=&N^{12}+6 N^{11}-27 N^{10}-186 N^9-197 N^8+310 N^7+899 N^6+1198 N^5+1020 N^4
\nonumber\\ &&
+112 N^3-192 N^2+64 N+64,
\\
P_{76} &=& 699 N^{12}+4194 N^{11}+16447 N^{10}+43214 N^9+42657 N^8-19098 N^7-36963 N^6
\nonumber\\ &&
-11670 N^5-45064 N^4-39392 N^3+7536 N^2+8064 N+1728,
\\
P_{77} &=& 723 N^{12}+4338 N^{11}+12623 N^{10}+17230 N^9-8583 N^8-30018 N^7+47709 N^6
\nonumber\\ &&
+75738 N^5+8776 N^4+67208 N^3+4416 N^2-41184 N-20736.
\end{eqnarray}
%----------------------------------------------------------------------------------------------------------------------------

%\end{eqnarray}
%--------------------------------------------------------------------------------------------------------
We mention the moment--relation 
%-----------------------------------------------------------------------------------------------
\begin{eqnarray}
\Delta \gamma_{gg}^{(2)}(N=1) = - 2 \beta_2,
\end{eqnarray}
%--------------------------------------------------------------------------------------------------------
with \cite{BETA,Chetyrkin:2017bjc}
%-----------------------------------------------------------------------------------------------
\begin{eqnarray}
\beta_2 = \frac{2857}{54} \textcolor{blue}{C_A^3}
+\textcolor{blue}{T_F N_F} \Biggl[
        -\frac{1415}{27} \textcolor{blue}{C_A^2}
        - \frac{205}{9} \textcolor{blue}{C_A C_F}
        +2 \textcolor{blue}{C_F^2}
\Biggr]
+\textcolor{blue}{(T_F N_F)^2} \Biggl[
        \frac{158}{27} \textcolor{blue}{C_A}
        +\frac{44}{9} \textcolor{blue}{C_F}
\Biggr].
\end{eqnarray}
%--------------------------------------------------------------------------------------------------------
The so-called supersymmetric relation obtained by setting $C_F = C_A = 2 T_F N_F$ 
is broken using dimensional regularization from $O(a_s^2)$ onward, while the 
relation 
%--------------------------------------------------------------------------------------------------------
\begin{eqnarray}
\label{eq:susy}
\Delta \gamma_{qq}^{(k)} +
\Delta \gamma_{gq}^{(k)} - 
\Delta \gamma_{qg}^{(k)} -
\Delta \gamma_{gg}^{(k)} = 0
\end{eqnarray}
%--------------------------------------------------------------------------------------------------------
also holds at $k=1$ using dimensional reduction \cite{Mertig:1995ny} in this 
limit.

Furthermore, one has
%--------------------------------------------------------------------------------------------------------
\begin{eqnarray}
\label{eq:qgsum}
\Delta \gamma_{qg}^{(k)}(N=1) = 0,~~\text{for}~k = 0, 1, 2.
\end{eqnarray}
%--------------------------------------------------------------------------------------------------------
The expansions have been performed using the {\tt HarmonicSums} command {\tt HarmonicSumsSeries} since
(\ref{eq:qg2},\ref{eq:gg2}) contain evanescent poles at $N=1$.
More generally, also the first moment of the gluonic Wilson coefficient, related to (\ref{eq:qgsum}), for the structure 
function  $g_1(x,Q^2)$ both for the massless \cite{Zijlstra:1993sh,FORTR} and the massive case in the asymptotic 
representation to two--loop order \cite{BIER15} vanishes. This is known at one--loop order even for general kinematics 
\cite{Watson:1981ce}.

The splitting functions in $z$--space are obtained by a Mellin inversion, cf.~(\ref{eq:INVmel}), and are given in
computer readable form in an attachment to the paper.
%--------------------------------------------------------------------------------------------------------
\section{Comparison to the literature}
\label{sec:5}
%--------------------------------------------------------------------------------------------------------

\vspace*{1mm}
\noindent
We confirm the results for the singlet anomalous dimensions calculated in \cite{Moch:2014sna},
where the on--shell forward Compton amplitude has been used for the computation. The contributions 
$\propto T_F$ have already been calculated independently as a by--product of the massive on--shell 
operator matrix elements in Ref.~\cite{Behring:2019tus}, 
to which we also agree. 
The comparison to the large $N_F$ expansions of Refs.~\cite{Gracey:1996ad,Bennett:1998sr} has been given in
our previous paper \cite{Behring:2019tus} already, where all these terms are covered.

Let us finally consider the small $z$ limit\footnote{For a review on the small $z$ predictions of
the different evolution kernels see \cite{Blumlein:1999ev}.} of the present results and compare to the predictions given in
\cite{Bartels:1996wc,Blumlein:1996hb}. In Mellin--$N$ space these terms are given by the most singular 
contribution expanding around $N=0$. One obtains at one-- and two--loop order
%--------------------------------------------------------------------------------------------------------
\begin{eqnarray}
\Delta \gamma^{(0), N \rightarrow 0} &=& -\frac{4}{N}\left(\begin{array}{cc}
C_F   & - 2 T_F N_F \\
2 C_F & 4 C_A 
\end{array} \right)
\\
\Delta \gamma^{(1), N \rightarrow 0} &=& -\frac{8}{N^3}\left(\begin{array}{cc}
C_F (2 C_A - 3 C_F - 4 T_F N_F) & 2 C_F (2 C_A + C_F) \\
- 2 C_F (2 C_A + C_F)            & 4 (2 C_A^2 - C_F T_F N_F) 
\end{array} \right)
\end{eqnarray}
%--------------------------------------------------------------------------------------------------------
which agree with the evolution kernels given in Refs.~\cite{Bartels:1996wc,Blumlein:1996hb}.

At three-loop order we have in the M--scheme
%--------------------------------------------------------------------------------------------------------
\renewcommand{\arraystretch}{1.5}
\begin{align}
& \Delta \gamma^{(2), N \rightarrow 0} = -\frac{16}{N^5}
\nonumber\\
& {\footnotesize
\times \left(\begin{array}{cc}
C_F (3 C_A^2 - 12 C_A C_F + 10 C_F^2 + T_F N_F( 12 C_A + 16 C_F)) &
2 T_F N_F (15 C_A^2 + 4 C_A C_F - 8 C_F T_F N_F)
\\
- 2 C_F (15 C_A^2 + 8 C_A C_F - 4 C_F^2 - 8 C_F T_F N_F) &
-4 (14 C_A^3 + T_F N_F(C_A^2  - 12 C_A C_F  - 2 C_F^2 ))
\end{array} \right)}
\end{align}
\renewcommand{\arraystretch}{1.0}
%--------------------------------------------------------------------------------------------------------

\noindent
and find a deviation both in case of $\Delta \gamma_{qg}^{(2)}$ and $\Delta \gamma_{gq}^{(2)}$, already noticed in
\cite{Vogt:2008yw}. The relative deviation amounts to $\sim \pm 3.2\%$, with the difference of the expression
in the M--scheme and the result obtained for the infrared evolution equation (IEE),
$\delta \Delta \gamma_{ij}^{(2), N \rightarrow 0}  = 
 \Delta \gamma_{ij}^{(2), N \rightarrow 0, \rm M}  - 
 \Delta \gamma_{ij}^{(2), N \rightarrow 0, \rm IEE}$,
%--------------------------------------------------------------------------------------------------------
\begin{eqnarray}
\delta \Delta \gamma_{qg}^{(2), N \rightarrow 0} &=& \frac{48}{N^5} C_F T_F N_F (C_A - C_F), 
\\
\delta \Delta \gamma_{gq}^{(2), N \rightarrow 0} &=& \frac{64}{N^5} C_F^2 (C_A - C_F).
\end{eqnarray}
%--------------------------------------------------------------------------------------------------------
One may consider a different theory but QCD by setting $C_A = C_F$. In this case the prediction 
\cite{Bartels:1996wc,Blumlein:1996hb} for the evolution kernels agrees with the perturbative calculation of  
the anomalous dimensions.

It has been the group of  J.~Kodaira \cite{Kiyo:1996si}, who also considered the effective Wilson coefficient in the 
non--singlet case Ref.~\cite{Bartels:1995iu}\footnote{In using infrared evolution equations the authors of 
Refs.~\cite{Kirschner:1983di,Bartels:1995iu,Bartels:1996wc} do not specify the factorization scheme, which complicates 
comparisons with calculations performed e.g. in the M--scheme.}, finding that these contributions 
are suppressed by a further power in $N$. This also applies to the effective Wilson coefficients in the 
non--singlet case and 
therefore the evolution kernels of \cite{Kirschner:1983di,Bartels:1995iu} agree\footnote{After correcting  
Ref.~\cite{Kirschner:1983di} in \cite{Blumlein:1995jp}.}. 
By considering the expansion of
the function to be interpreted as a matrix formulation for the effective Wilson coefficient in \cite{Bartels:1996wc},
%--------------------------------------------------------------------------------------------------------
\begin{eqnarray}
\frac{N}{{\bf 1} N - \tfrac{1}{8 \pi^2} {\bf F}_0}  = {\bf 1} + \sum_{k = 1}^\infty \left(\frac{a_s}{N^2}\right)^k {\bf 
F}_{0,k}.
\end{eqnarray}
%--------------------------------------------------------------------------------------------------------
The expansion coefficients ${\bf F}_{0,k}$ are $2 \times 2$ matrices only depending on color factors.
Therefore, in the representation of \cite{Bartels:1996wc}, the Wilson coefficients are suppressed
by one power in $N$, like in the non--singlet case, cf.~\cite{Bartels:1995iu,Kiyo:1996si}. 
To perform a full comparison one has to form two scheme invariant quantities. This is not really possible 
in pure polarized QCD, since there
is only one structure function $g_1(x,Q^2)$ and also considering the Wilson coefficients 
\cite{Zijlstra:1993sh,FORTR}.\footnote{At the level of twist--2 the structure function $g_2(x,Q^2)$
is not an independent quantity, but related by the Wandzura--Wilczek relation \cite{Wandzura:1977qf}
to the structure function $g_1(x,Q^2)$.
One might consider the physical pair $\{g_1(x,Q^2), \partial g_1(x,Q^2)/\partial \ln(Q^2)\}$, cf.~\cite{G1pair}.
However, the so--called leading powers in $N$ are here not of the same order.} 
In \cite{Moch:2014sna} scheme invariant polarized evolution kernels
for the contributions to the structure function $g_1$ and additional fictitious gravitational contributions in the 
gluonic channels have been formed for which the prediction in \cite{Bartels:1996wc,Blumlein:1996hb} hold. 

In data analyses or the phenomenological description of the polarized structure functions the consideration of only the 
leading small $z$ terms in the evolution kernels is numerically not sufficient. Subleading terms dominate over the 
leading order effects, cf.~\cite{Blumlein:1995jp,Blumlein:1996hb,Blumlein:1999ev}.
%-------------------------------------------------------------------------------------------------------- 
\section{Conclusions} 
\label{sec:6} 
%--------------------------------------------------------------------------------------------------------

\vspace*{1mm}
\noindent
We have calculated the polarized three--loop singlet anomalous dimensions $\Delta \gamma_{qq}^{(2),\rm PS}$, 
$\Delta \gamma_{qg}^{(2)}$, $\Delta \gamma_{gq}^{(2)}$, $\Delta \gamma_{gg}^{(2)}$, 
and the non--singlet anomalous 
dimension $\Delta \gamma_{qq}^{(2), \rm s, NS}$ in Quantum Chromodynamics and agree with the results of 
Refs.~\cite{Moch:2014sna,Moch:2015usa,Behring:2019tus}. The singlet anomalous dimensions have been calculated by using 
the method of massless off--shell OMEs, which has been applied for this purpose for the first time.
The calculation has been fully automated referring to the Larin scheme, performing a finite transformation to the 
M--scheme for the final results. Both schemes are valid to describe the scaling violations of the polarized structure 
functions, however, with a different outcome for the polarized parton distribution functions, which are scheme--dependent
quantities. Comparing to predictions of the leading small $z$ behaviour one finds deviations for the off--diagonal
elements at three--loop order, from which one concludes 
that the calculation in Ref.~\cite{Bartels:1996wc}
is not in the M--scheme starting with three--loop order. To obtain the complete picture, one has to consider also the 
associated behaviour of the Wilson coefficients and to form scheme--invariant quantities. We also mention that partial 
checks on the polarized anomalous 
dimensions are possible from the pole terms of the single- and two--mass massive OMEs to three loop order, 
cf.~\cite{MASSIVEome}. Both the anomalous dimension and splitting functions are given in computer readable form in 
the file {\tt ANOM3pol.m} attached to this paper.

\vspace{5mm}\noindent 
{\bf Acknowledgment.}~We would like to thank J.~Ablinger and A.~Vogt for a discussion. This project has 
received funding from the European Union's Horizon 2020 research and innovation programme under the Marie 
Sk\l{}odowska--Curie grant agreement No. 764850, SAGEX and from the Austrian Science Fund (FWF) grants 
SFB F50 (F5009-N15) and P33530.

\appendix
%-------------------------------------------------------------------------------------
\section{A new polarized gluonic Feynman rule}
\label{sec:A}
%-------------------------------------------------------------------------------------

\vspace*{1mm}
\noindent
The gluonic OME $\Delta A_{gg}$ requires a new Feynman rule to 
calculate the anomalous dimensions in the massless case, containing
a local operator with with five external gluon lines, extending the
setting given in Refs.~\cite{Bierenbaum:2009mv, Behring:2019tus}.
The operator insertion reads in the polarized case
%-------------------------------------------------------------------------------------
\begin{eqnarray}
  \raisebox{-25pt}{\includegraphics[width=0.3\textwidth]{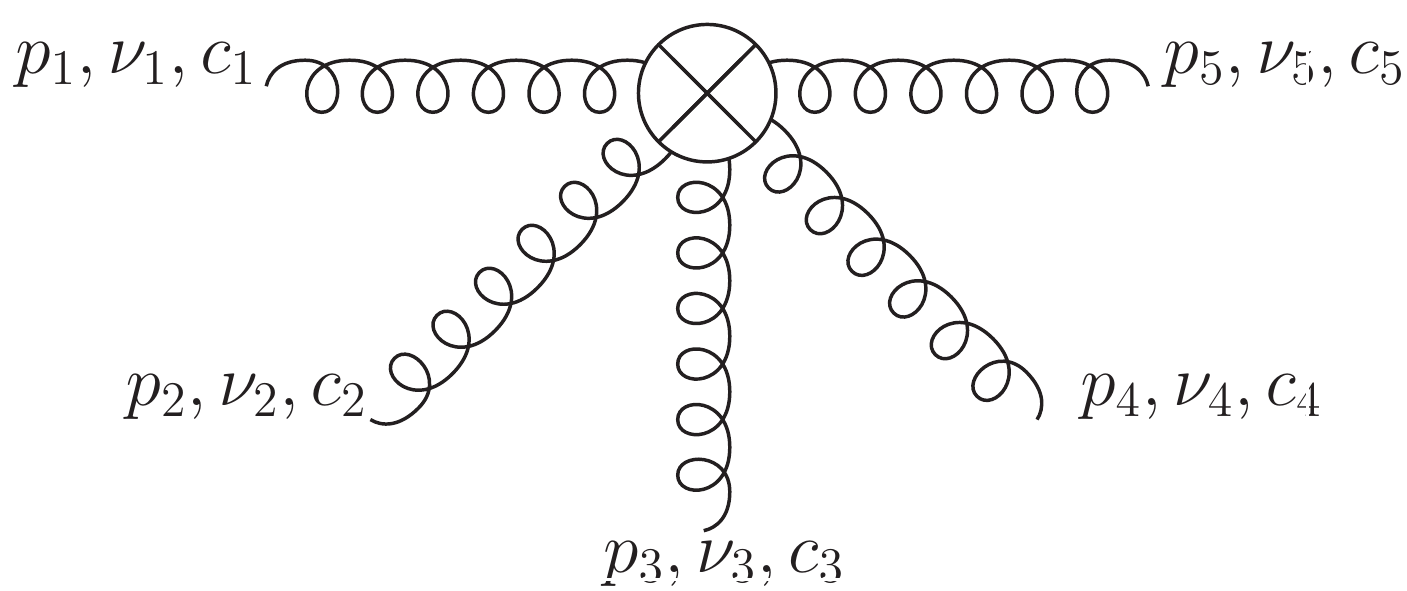}}
  &=& -i g^3 \frac{1-(-1)^N}{2}
	\Biggl( 
    \nonumber \\  &&
\hspace*{-1.6cm}
\hphantom{+}  f^{c_{1}c_{2}c_{x}} f^{c_{3}c_{x}c_{y}} f^{c_{4}c_{5}c_{y}} O_{\nu_{1}\nu_{2}\nu_{3}\nu_{4}\nu_{5}}
(p_{1},p_{2},p_{3},p_{4},p_{5}) 
\nonumber \\ && 
\hspace*{-6.1cm}
+ f^{c_{1}c_{2}c_{x}} f^{c_{4}c_{x}c_{y}} f^{c_{3}c_{5}c_{y}} 
O_{\nu_{1}\nu_{2}\nu_{4}\nu_{3}\nu_{5}}(p_{1},p_{2},p_{4},p_{3},p_{5}) 
%\nonumber \\ && 
+ f^{c_{1}c_{2}c_{x}} f^{c_{5}c_{x}c_{y}} f^{c_{3}c_{4}c_{y}} 
O_{\nu_{1}\nu_{2}\nu_{5}\nu_{3}\nu_{4}}(p_{1},p_{2},p_{5},p_{3},p_{4}) \nonumber \\ && 
\hspace*{-6.1cm}
+ f^{c_{1}c_{3}c_{x}} f^{c_{2}c_{x}c_{y}} f^{c_{4}c_{5}c_{y}} 
O_{\nu_{1}\nu_{3}\nu_{2}\nu_{4}\nu_{5}}(p_{1},p_{3},p_{2},p_{4},p_{5}) 
%\nonumber \\ && 
+ f^{c_{1}c_{3}c_{x}} f^{c_{4}c_{x}c_{y}} f^{c_{2}c_{5}c_{y}} 
O_{\nu_{1}\nu_{3}\nu_{4}\nu_{2}\nu_{5}}(p_{1},p_{3},p_{4},p_{2},p_{5}) \nonumber \\ && 
\hspace*{-6.1cm}
+ f^{c_{1}c_{3}c_{x}} f^{c_{5}c_{x}c_{y}} f^{c_{2}c_{4}c_{y}} 
O_{\nu_{1}\nu_{3}\nu_{5}\nu_{2}\nu_{4}}(p_{1},p_{3},p_{5},p_{2},p_{4}) 
%\nonumber \\ && 
+ f^{c_{1}c_{4}c_{x}} f^{c_{2}c_{x}c_{y}} f^{c_{3}c_{5}c_{y}} 
O_{\nu_{1}\nu_{4}\nu_{2}\nu_{3}\nu_{5}}(p_{1},p_{4},p_{2},p_{3},p_{5}) 
\nonumber \\ && 
\hspace*{-6.1cm}
+ f^{c_{1}c_{4}c_{x}} f^{c_{3}c_{x}c_{y}} f^{c_{2}c_{5}c_{y}} 
O_{\nu_{1}\nu_{4}\nu_{3}\nu_{2}\nu_{5}}(p_{1},p_{4},p_{3},p_{2},p_{5}) 
%\nonumber \\ && 
+ f^{c_{1}c_{4}c_{x}} f^{c_{5}c_{x}c_{y}} f^{c_{2}c_{3}c_{y}} 
O_{\nu_{1}\nu_{4}\nu_{5}\nu_{2}\nu_{3}}(p_{1},p_{4},p_{5},p_{2},p_{3}) \nonumber \\ && 
\hspace*{-6.1cm}
+ f^{c_{1}c_{5}c_{x}} f^{c_{2}c_{x}c_{y}} f^{c_{3}c_{4}c_{y}} 
O_{\nu_{1}\nu_{5}\nu_{2}\nu_{3}\nu_{4}}(p_{1},p_{5},p_{2},p_{3},p_{4}) %\nonumber \\ && 
+ f^{c_{1}c_{5}c_{x}} f^{c_{3}c_{x}c_{y}} f^{c_{2}c_{4}c_{y}} 
O_{\nu_{1}\nu_{5}\nu_{3}\nu_{2}\nu_{4}}(p_{1},p_{5},p_{3},p_{2},p_{4}) \nonumber \\ && 
\hspace*{-6.1cm}
+ f^{c_{1}c_{5}c_{x}} f^{c_{4}c_{x}c_{y}} f^{c_{2}c_{3}c_{y}} 
O_{\nu_{1}\nu_{5}\nu_{4}\nu_{2}\nu_{3}}(p_{1},p_{5},p_{4},p_{2},p_{3}) %\nonumber \\ && 
+ f^{c_{2}c_{3}c_{x}} f^{c_{1}c_{x}c_{y}} f^{c_{4}c_{5}c_{y}} 
O_{\nu_{2}\nu_{3}\nu_{1}\nu_{4}\nu_{5}}(p_{2},p_{3},p_{1},p_{4},p_{5}) \nonumber \\ && 
\hspace*{-6.1cm}
+ f^{c_{2}c_{4}c_{x}} f^{c_{1}c_{x}c_{y}} f^{c_{3}c_{5}c_{y}} 
O_{\nu_{2}\nu_{4}\nu_{1}\nu_{3}\nu_{5}}(p_{2},p_{4},p_{1},p_{3},p_{5}) %\nonumber \\ && 
+ f^{c_{2}c_{5}c_{x}} f^{c_{1}c_{x}c_{y}} f^{c_{3}c_{4}c_{y}} 
O_{\nu_{2}\nu_{5}\nu_{1}\nu_{3}\nu_{4}}(p_{2},p_{5},p_{1},p_{3},p_{4}) 
	\Biggr)
\nonumber\\
\end{eqnarray}
%-------------------------------------------------------------------------------------
with 
%-------------------------------------------------------------------------------------
\begin{eqnarray}
	\lefteqn{O_{\nu_1 \nu_2 \nu_3 \nu_4 \nu_5}(p_1,p_2,p_3,p_4,p_5) = } 
	\nonumber \\
	&&
\Delta_{\nu_3}
	\biggl[
		  \Delta_{\nu_5} \varepsilon_{\Delta \nu_1 \nu_2 \nu_4}
		- \Delta_{\nu_4} \varepsilon_{\Delta \nu_1 \nu_2 \nu_5}
	\biggr] 
	\sum\limits_{m=2}^{N-1} 
	\left( \Delta.p_1 + \Delta.p_2 \right)^{m-2} 
	\left( - \Delta.p_4 - \Delta.p_5 \right)^{N-m-1}
	\nonumber \\ &&
	+ \tilde{O}_{\nu_1 \nu_2 \nu_3 \nu_4 \nu_5}(p_1,p_2,p_3,p_4,p_5)
	+ \tilde{O}_{\nu_2 \nu_1 \nu_3 \nu_5 \nu_4}(p_2,p_1,p_3,p_5,p_4)
	+ \tilde{O}_{\nu_4 \nu_5 \nu_3 \nu_2 \nu_1}(p_4,p_5,p_3,p_2,p_1)
	\nonumber \\ &&
	+ \tilde{O}_{\nu_5 \nu_4 \nu_3 \nu_2 \nu_1}(p_5,p_4,p_3,p_2,p_1)
\end{eqnarray}
%-------------------------------------------------------------------------------------
and
%-------------------------------------------------------------------------------------
\begin{eqnarray}
	\lefteqn{\tilde{O}_{\nu_1 \nu_2 \nu_3 \nu_4 \nu_5}(p_1,p_2,p_3,p_4,p_5) = } 
	\nonumber \\ && 
	-\Delta_{\nu_2} \Delta_{\nu_3} \Delta_{\nu_4}
	\biggl[
		\Delta_{\nu_5} \varepsilon_{\Delta p_5 p_1 \nu_1}	
		+ p_5.\Delta \varepsilon_{\Delta \nu_5 \nu_1 p_1}	
	\biggr]
	\sum \limits_{m=2}^{N-3}
	\sum\limits_{n=m+1}^{N-2}
	\sum\limits_{o=n+1}^{N-1}
	\left( \Delta.p_1 \right)^{m-2}
	\left( \Delta.p_1 + \Delta.p_2 \right)^{n-m-1}
	\nonumber \\ && \times
	\left( -\Delta.p_4 - \Delta.p_5 \right)^{o-n-1}
	\left( - \Delta.p_5 \right)^{N-o-1}
	+ \Delta_{\nu_3} \Delta_{\nu_4}
	\biggl[
		  \Delta_{\nu_2} \varepsilon_{\Delta p_5 \nu_1 \nu_5}	
		- \Delta_{\nu_1} \varepsilon_{\Delta p_5 \nu_2 \nu_5}	
	\biggr]
	\nonumber \\ && \times
	\sum \limits_{m=2}^{N-2}
	\sum\limits_{n=m+1}^{N-1}
	\left( \Delta.p_1 + \Delta.p_2 \right)^{m-2}
	\left( -\Delta.p_4 - \Delta.p_5 \right)^{n-m-1}
	\left( - \Delta.p_5 \right)^{N-n-1}.
\end{eqnarray}
%-------------------------------------------------------------------------------------
All momenta are inflowing and the symbols $f^{abc}$ denote the structure constants of
$SU(N_c)$. For the sums in the Feynman rule it is understood that the upper summation bound 
is larger or equal than the lower bound.
%-------------------------------------------------------------------------------------

{%\footnotesize
%-------------------------------------------------------------------------------------

}
%-----------------------------------------------------------------------------------------------------
\end{document}